\documentclass[12pt]{iopart}

\usepackage[english]{babel}
\usepackage[T1]{fontenc}

\usepackage{lmodern}

\usepackage{iopams}
\usepackage{graphicx}
\usepackage{pstricks}
\usepackage{pst-node}
\usepackage{amsmath}
\usepackage[normalem]{ulem}

\usepackage{cite}      
\usepackage{graphicx}  
\usepackage{amsmath}   
\usepackage{url}
\usepackage{blkarray}
\usepackage{multirow}
\usepackage{pstricks}
\usepackage{pst-node}
\usepackage{multirow}
\usepackage{rotating}
\usepackage{amsfonts}
\usepackage{amssymb}
\usepackage{exscale}
\usepackage{upgreek}
\usepackage{textcomp}
\usepackage{epsfig}
\usepackage[T1]{fontenc}
\usepackage{perpage}

\usepackage{gensymb}

\usepackage{subfig}

\usepackage{siunitx} 

\usepackage{amsthm}

\begin{document}

\newcommand{\TM}{{\mathrm{T}\Omega}}
\newcommand{\TxM}{{\mathrm{T}_x\Omega}}
\newcommand{\TstarM}{{\mathrm{T}^{*}\Omega}}
\newcommand{\TstarxM}{{\mathrm{T}^{*}_x\Omega}}

\theoremstyle{definition} 
\newtheorem{mydef}{Definition}
\newtheorem{mytheorem}{Theorem}
\newtheorem{myremark}{Remark}
\newtheorem{myrp}{RP}
\newtheorem{myconjecture}{Conjecture}
\newtheorem{myproblem}{Inverse problem}

\title[]{Effects of device geometry and material properties on dielectric losses in superconducting coplanar-waveguide resonators}

\author{V. Lahtinen$^1$ and M. M\"ott\"onen$^{1,2}$}

\address{$^1$QCD Labs, QTF Centre of Excellence, Department of Applied Physics, Aalto University, FI-00076 Aalto, Finland \\
$^2$QTF Centre of Excellence, VTT Technical Research Centre of Finland, P.O. Box 1000, FI-02044 VTT, Finland}

\ead{valtteri.lahtinen@aalto.fi}

\begin{abstract}
Superconducting coplanar-waveguide (CPW) resonators are one of the key devices in circuit quantum electrodynamics (cQED). Their performance can be limited by dielectric losses in the substrate and in the material interfaces. Reliable modeling is required to aid in the design of low-loss CPW structures for cQED. We analyze the geometric dependence of the dielectric losses in CPW structures using finite-element modeling of the participation ratios of the lossy regions. 
In a practical scenario, uncertainties in the the dielectric constants and loss tangents of these regions introduce uncertainties in the theoretically predicted participation ratios.
We present a method for combining loss simulations with measurements of two-level-system-limited quality factors and resonance frequencies of CPW resonators. Namely, we solve an inverse problem to find model parameters producing the measured values. High quality factors are obtainable by properly designing the cross-sectional geometries of the CPW structures, but more accurate modeling and design methods for low-loss CPW resonators are called for major future improvements. Our nonlinear optimization methodology for solving the aforementioned inverse problem is a step in this direction.

\end{abstract}


\maketitle

\section{Introduction}

Defects that can be modeled as two-level systems (TLSs) are a central source of loss in circuit quantum electrodynamics (cQED), in devices such as superconducting coplanar-waveguide (CPW) resonators and qubits \cite{Wenner, Wang09, Barends, Sage, Martinis, Geerlings, Megrant, Quintana, Calusine}. The interactions between TLS defects and the electric field present in dielectric materials give rise to losses, consequently reducing resonator quality factors and qubit lifetimes. In particular, the performance of superconducting CPW resonators, essential in cQED \cite{Wallraff, Blais}, is thus limited by dielectric losses in the substrate and in the material interfaces, where thin layers of amorphous materials containing TLS defects are present. In terms of electromagnetic modeling, these losses are manifested in the complex permittivity of the material. Reliable numerical modeling and reliable material property data are required to mitigate losses in such devices.  

Dielectric losses in CPW resonators and qubits have recently been studied in several works \cite{Wenner, O'Connell, Wang, Dial, Gambetta, Calusine, Woods}. These rely on a common approach for estimating the TLS-limited quality factor of a resonator, $Q_\mathrm{TLS}$, by computing for each lossy region in the device the \emph{participation ratio}, i.e., the ratio of the electric field energy in the region to the total electric field energy in the resonator. The amount of losses induced by a given region is directly proportional to its participation ratio. Utilizing this approach, several different methodologies have been used to investigate the sources of these losses. Both two-dimensional (2D) \cite{Wenner, Calusine, Woods} and three-dimensional (3D) \cite{Dial, Niepce} simulations have been utilized extensively both in electrostatic \cite{Wenner, Calusine, Woods} and high-frequency formulations \cite{Gambetta, Dial}. Recently, in reference~\cite{Niepce}, Niepce~{\it et al.} combined participation ratio simulations with 3D simulations of Maxwell-London equations to take also into account the magnetic-field penetration in a disordered superconductor. Wang~{\it et al.} on the other hand, combined local electrostatic and global high-frequency simulations to study dielectric loss in qubits~\cite{Wang}.

For computing $Q_\mathrm{TLS}$ from participation ratio simulations, one also needs information about the \emph{loss tangents} of the lossy regions. One option is to simply assume some literature values for the loss tangents and predict the resulting quality factors accordingly \cite{Wenner}. In contrast, Calusine {\it et al.}~\cite{Calusine} presented a method for finding the loss tangents by combining participation ratio simulations with quality factor measurements. Although etching into the substrate has been shown to be an effective way to mitigate dielectric losses~\cite{Wenner, Calusine, Vissers, Bruno}, for typical anisotropically etched structures, the relative changes in the interface participation ratios are almost equal with each other for changes in the trench depth, which makes it difficult to differentiate between the different interfaces using such data. This results in high uncertainties in the solved loss tangents even for rather small uncertainty in the input values.
In reference~\cite{Woods}, isotropic etching was utilized to produce a set of CPW structures, resulting in participation ratio data with more distinguishable interfaces, and thus a more reliable loss tangent prediction.

The participation ratios depend on the dielectric constants and on the thicknesses of the lossy interfaces. By considering infinitesimally thin interface layers, the electromagnetic boundary conditions imply a simple dependence between these values and participation ratios~\cite{Wenner}. With some further assumptions, the simulated losses for certain values for the dielectric constant and the thickness may be scaled for all other values accordingly in a simple manner~\cite{Calusine, Woods}.

In this paper, we investigate the significant geometric features of CPW structures affecting the dielectric losses by carrying out 2D electrostatic finite-element simulations to compute the participation ratios of the lossy regions. Furthermore, we study how the dielectric constants of the lossy regions affect the resulting participation ratios, without making further simplifying assumptions, in addition to those arising from the discretization, about the field profile in the interfacial layers. In particular, our simulations indicate that the above-discussed proportionality relationships connecting participation ratios and dielectric constants do not hold in all cases for all interfacial regions. This serves as a motivation for developing a method to solve the dielectric constants of the lossy regions from resonance frequency measurement data. Combining quality factor measurement data with the resonance frequency data, we formulate an inverse problem to solve for the loss tangents and the dielectric constants. This is in contrast to recent works, where either feasible values for loss tangents and dielectric constants have been assumed \cite{Wenner}, or approximate formulas have been utilized to account for the dielectric constant values \cite{Calusine, Woods}. This is a step toward a more accurate design methodology for CPW resonators.

In section~2, we briefly present the theoretical approach to obtain the participation ratios. In section~3, we carry out simulations on CPW resonator cross-sections to study the influence the geometric and material-specific features have on the dielectric losses. In section~4, we discuss how to computationally determine the dielectric constants and loss tangents of the different regions in a set of CPW resonators, and utilize the presented approaches in section~5 for a case study of four CPW resonator cross-sections, demonstrating their feasibility. Finally, in section~6, we summarize this work and draw conclusions.

\section{Theoretical background}

We model the 2D cross-sections of CPW resonators, assuming a field profile of a long waveguide. 
Following reference~\cite{Calusine}, we solve the electrostatic problem in the scalar-potential formulation to find out the electric potential in the modeling domain $\Omega$. To obtain the field solutions, we utilize the commercial finite-element method (FEM) software \texttt{Comsol Multiphysics}~\cite{Comsol}.
To extract the dielectric losses, we compute the participation ratios, exhibited by each lossy dielectric region in the cross-section of a CPW resonator. In the case of weak dissipation, the participation ratio $p_i$ for a region $\Omega_i$ in the modeling domain $\Omega$ is defined as
\begin{equation}\label{p_i} 
p_i = \frac{ \frac{1}{2}\int_{\Omega_i} {\bf E} \cdot {\bf D} \rmd V}{\frac{1}{2}\int_{\Omega} {\bf E} \cdot {\bf D} \rmd V}, 
\end{equation}
where {\bf E} is the electric field intensity and {\bf D} is the electric displacement field obtained from the static problem.
That is, the participation ratio $p_i$ is the ratio of the electric field energy stored in the region $\Omega_i$ to the total electric field energy in $\Omega$. We assume linear and isotropic materials such that each material subdomain $\Omega_i$ is characterized by a real-valued dielectric constant $\epsilon_i$, i.e., the real part of the relative permittivity. Since our modeling domains consist merely of the CPW cross-sections, the volume integrals in equation~\eqref{p_i} reduce to surface integrals. Thus we obtain 
\begin{equation}\label{p_i2d} 
p_i = \frac{\frac{1}{2} \int_{\Omega_i} \epsilon_i  \|{\bf E}\|^2 \rmd A}{ \frac{1}{2} \int_{\Omega} \epsilon \|{\bf E}\|^2 \rmd A}, 
\end{equation}
where $\epsilon$ is the real-valued subdomain-wise-constant spatially dependent dielectric constant.

Along the lines of references~\cite{Wenner, Dial, Gambetta, Calusine}, knowing the participation ratios, one can compute the TLS-limited quality factor for a CPW resonator by
\begin{equation}\label{Q_TLS} 
\frac{1}{Q_{\mathrm{TLS}}} = \sum_i p_i \mathrm{tan}(\delta_i), 
\end{equation}
where $\mathrm{tan} (\delta_i)$ is the loss tangent associated with the region $\Omega_i$; the complex-valued nature of permittivity is thus taken into account by the real-valued dielectric constant and loss tangent together.

\section{Simulations of dielectric losses in CPW resonators}

In this section, we first investigate the dependence of dielectric losses on the geometric properties of the cross-section of a CPW resonator. Then, we move on to simulate the effects of variations in the dielectric constants of the lossy regions. Here, we pick a set of experimentally feasible parameters and study the effects of uncertainties in these parameters in the following sections.

We model the lossy dielectric interfaces as 5 nm thick flat regions at all the material interfaces: metal-to-air (MA), metal-to-substrate (MS) and substrate-to-air (SA). Because of these high-aspect-ratio structures, finite-element meshing needs to be carried out with extra care. To ensure the high quality of the mesh, we initially utilize a brief convergence analysis, refining the mesh until the relative changes in the participation ratios remain within approximately 1$\%$ from one mesh to another. With some geometry-to-geometry variation, our meshes result in roughly $10^6$ degrees of freedom to be solved for a given problem. 

\begin{figure}[!h]
	\centering

		\subfloat[][]{\includegraphics[width=7.5cm, height=4.55cm]{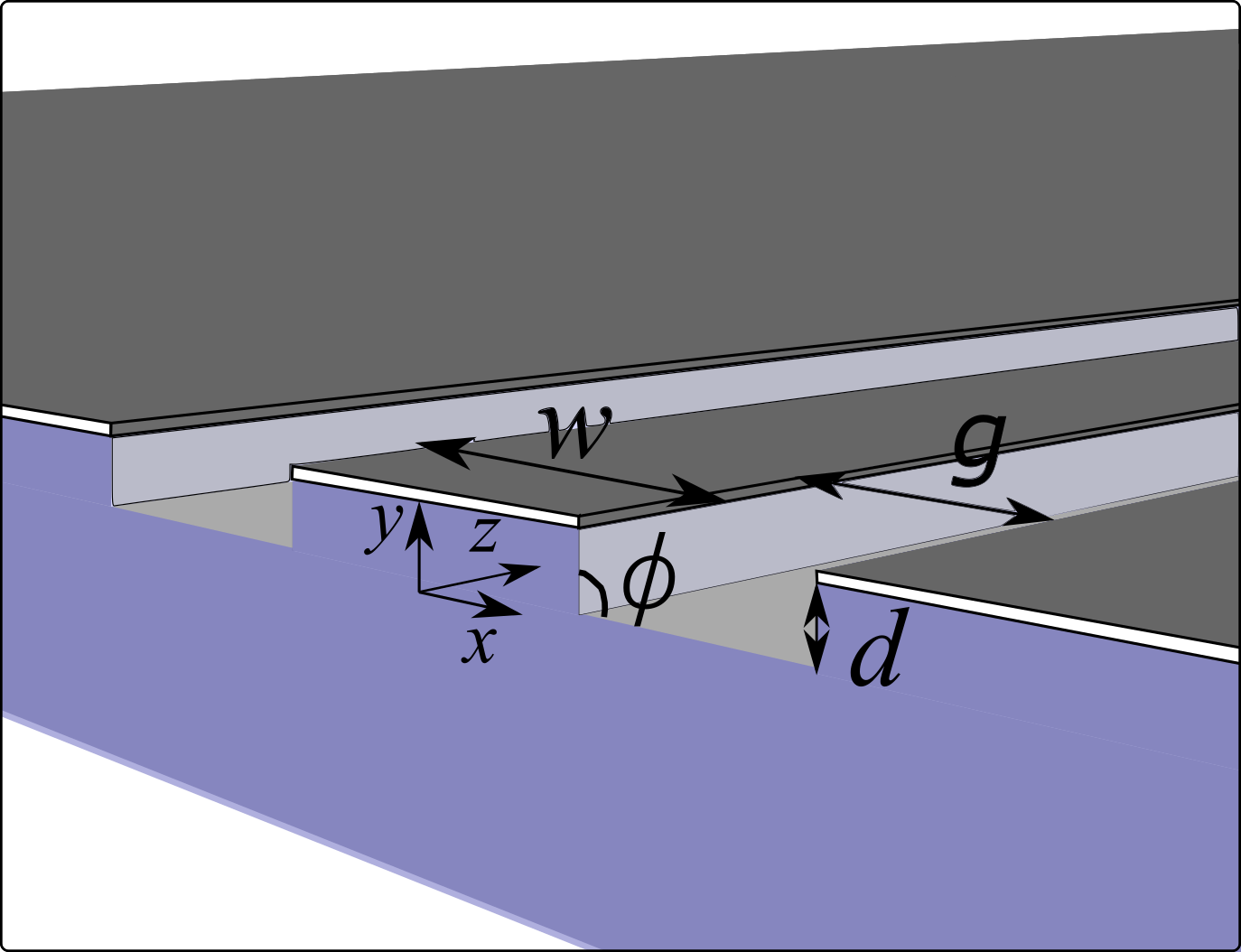}}  \\	
		\subfloat[][]{\includegraphics[width=7.5cm, height=5.0cm]{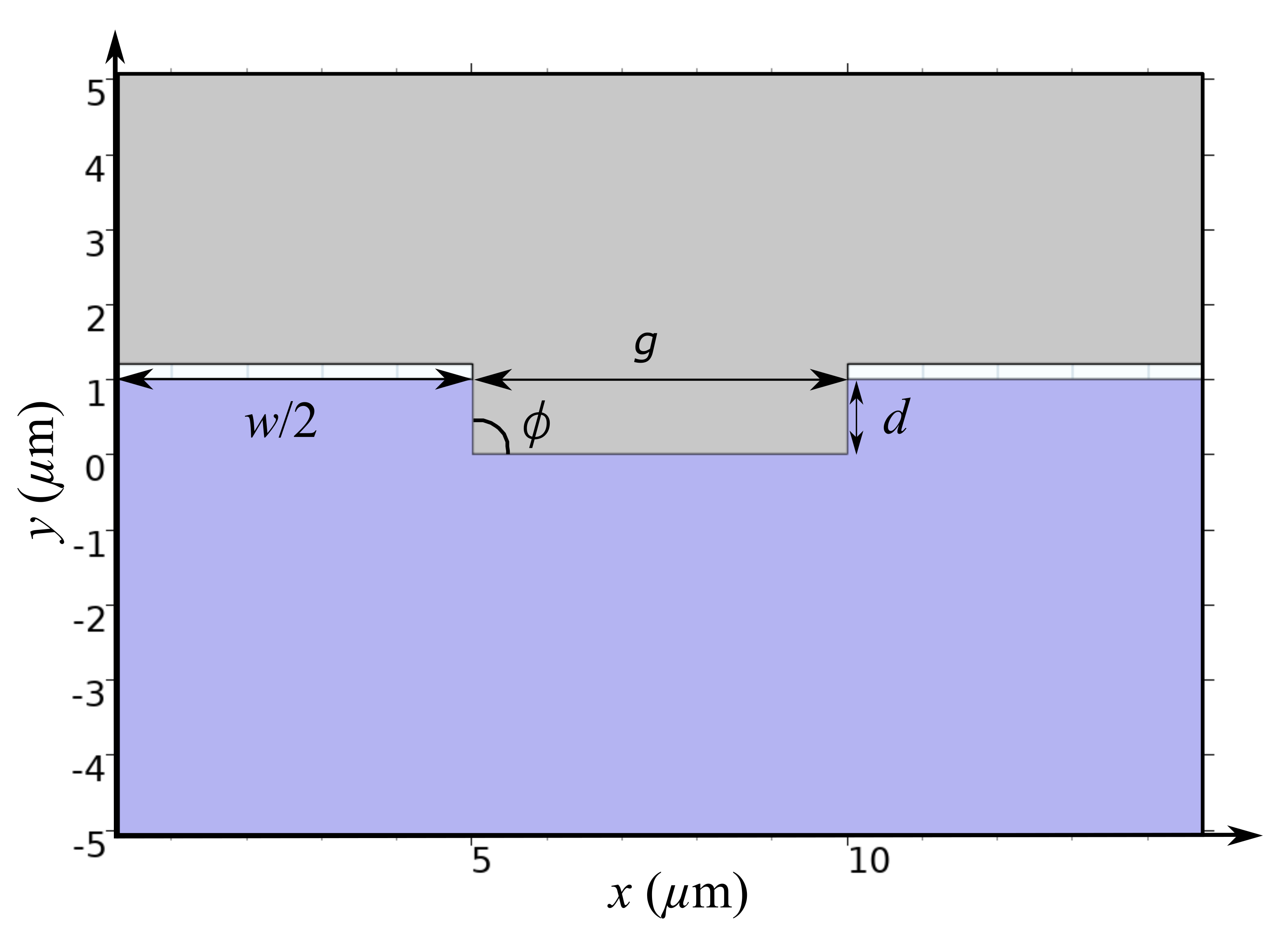}} 	\hspace{0.002cm} 
	\subfloat[][]{\includegraphics[scale=0.4525]{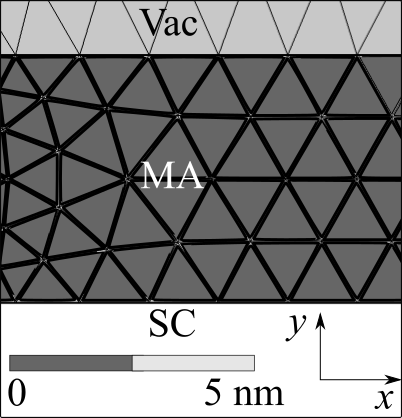}}

	\caption{
	(a) Three-dimensional view of a cut CPW resonator illustrating its cross-section and the definitions of the parameters $w$, $g$, $d$, and $\phi$. The figure is not to scale. (b) Part of our modeling domain for the cross-section of the CPW.
	The bottom part (blue color) contains the substrate, the top part the vacuum (grey color), and the white regions represent the superconductors (SC). Here, the trench depth $d = 1$~$\mu$m, sidewall angle $\phi = 90 \degree$,
	gap to ground $g = 5$ $\mu$m, and center conductor width $w = 10$~$\mu$m. For any sidewall angle, $w$, $g$, and $d$ are defined at the bottom surface of the superconductors as shown here. In particular, $d$ is the $y$-directional distance from the surface to the substrate at the bottom of the trenched gap, including the interfaces. This suits our purposes, even though it renders some ($w$,$g$,$d$,$\phi$) combinations impossible. Note also, that different sidewall angles result in different trench volumes. We simulate the cross-section for $x \geq 0$ only, utilizing symmetry. The height of the superconductors is $h = 200$~nm. We assume an identical sidewall angle for the superconductors as for the substrate. In principle, we assume an infinite ground plane, but for practical simulations, we cut off the modeling domain far away from the gap where fields are weak. At the symmetry boundary $x=0$ and at the outer boundaries (generally $x > w+2g$, $\vert y \vert > 4 d$, for this geometry $x = 30~\mu$m, $\vert y \vert = 20~\mu$m), we set ${\bf D} \cdot {\bf n} = 0$, where {\bf n} is the vector normal to the boundary.
	At the boundaries of the superconductors, we set Dirichlet boundary conditions and apply a potential difference between the center conductor and the ground plane of the CPW. (c) Mesh used in a simulation in the vicinity of a thin metal-to-air (MA) interface layer.}

	\label{fig:domain}
\end{figure}

In this section, we use a dielectric constant $\epsilon_i = 10$ and a loss tangent $\mathrm{tan} (\delta_i) = 0.002$ for all thin interface regions $\Omega_i$. These are rather typical values used in the literature \cite{Calusine,Dial,Wang,Wenner}. Utilizing equal dielectric constants and loss tangents for all interface layers helps us to extract and understand the trends in the participation ratios with varying geometric parameters, and compare them with the reported literature values. For the silicon substrate, we use $\epsilon_\mathrm{Si} = 11.6$ and $\mathrm{tan} (\delta_\mathrm{Si}) = 7.5 \times 10^{-7}$. \pagebreak
Note that the used values of the loss tangents are somewhat conservative.\footnote{See references~\cite{Wenner,Wang}. In reference~\cite{Calusine}, loss tangents of the order of $5 \times 10^{-4}$ are reported for the SA and MS interfaces and $10^{-7}$ for the substrate, leading correspondingly to higher $Q_\mathrm{TLS}$ values than estimated in the simulations of this section.} See figure~\ref{fig:domain} for a depiction of the modeling domain used in the simulations and an illustration of the mesh in a narrow region. Note that mesh quality in these narrow regions and corners is very important for reliable simulations. To ensure the accuracy, the meshes are structured so that equal mesh density is kept throughout the narrow interface regions, with finer regions at the corners. In the straight parts of the interface regions, the height of each element is approximately 1/4th of the thickness of the interface region, as shown in figure~\ref{fig:domain}. In the corners exhibiting a finer mesh, the size of the longest edge of the smallest element can be as small as 1/10th of the thickness of the interface region.

\subsection{Effect of the trench depth and sidewall angle}

\begin{figure}[!h]
	\centering 
	\subfloat[][]{\includegraphics[scale=0.9]{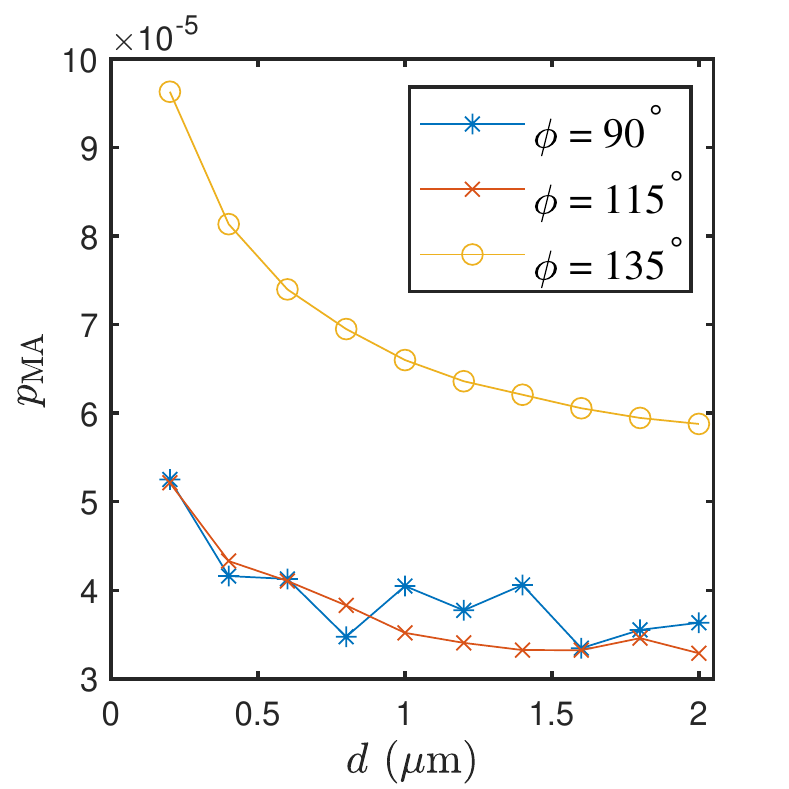}}
	\subfloat[][]{\includegraphics[scale=0.9]{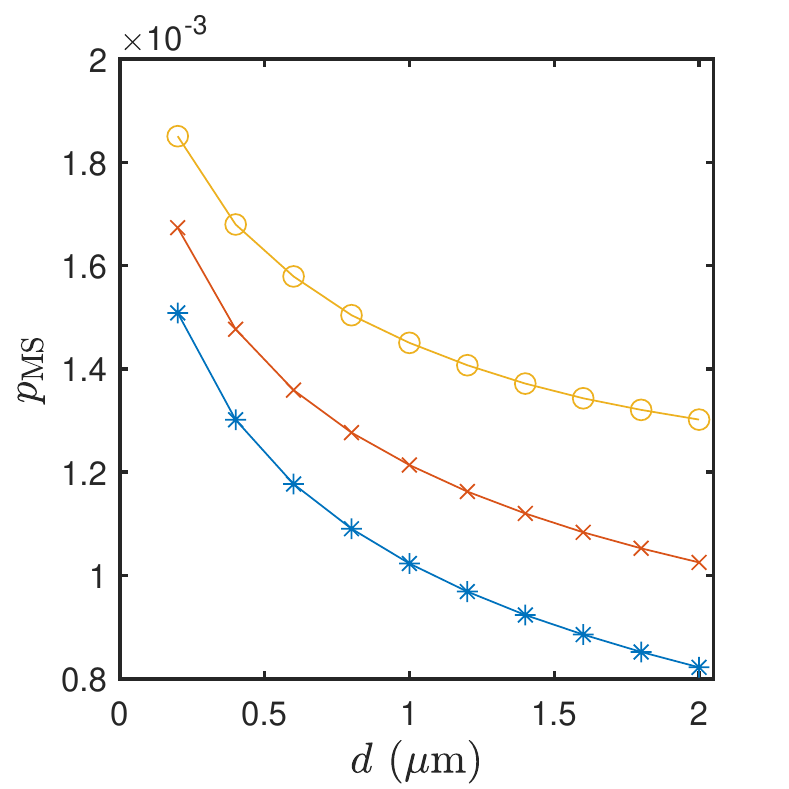}} \\
	\subfloat[][]{\includegraphics[scale=0.9]{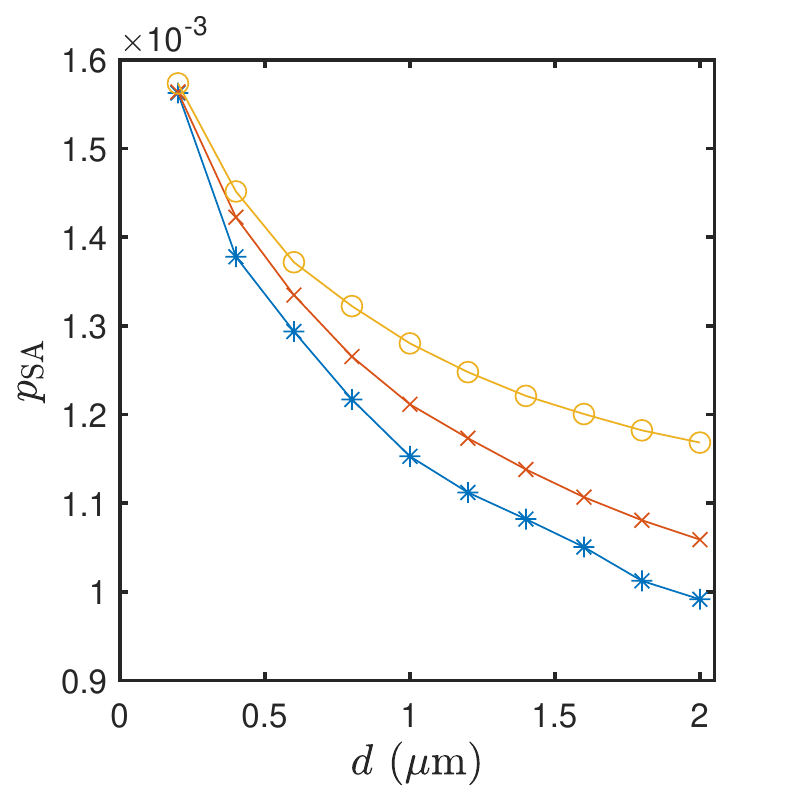}}
	\subfloat[][]{\includegraphics[scale=0.9]{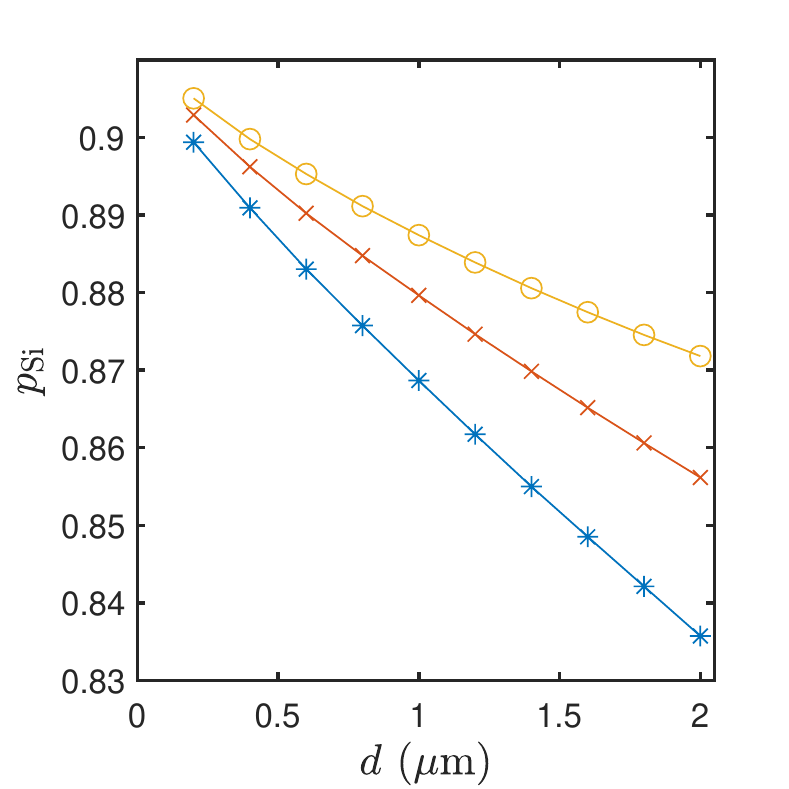}}
	
	\caption{Participation ratios (markers) of the (a) metal-to-air, (b) metal-to-substrate, and (c) substrate-to-air interfaces and (d) of the silicon substrate as functions of the trench depth $d$  for different sidewall angles $\phi$ as indicated. The solid lines are guides for the eye. Here, $w = 10 \textrm{ } \mu\textrm{m}$, $g = 5 \textrm{ } \mu\textrm{m}$, $\epsilon_\mathrm{SA} = \epsilon_\mathrm{MS} = \epsilon_\mathrm{MA} = 10$, and $\epsilon_\mathrm{Si} = 11.6$. 
	}\label{fig:p_i}
\end{figure}

\begin{figure}[!h]
	\centering
	\includegraphics[]{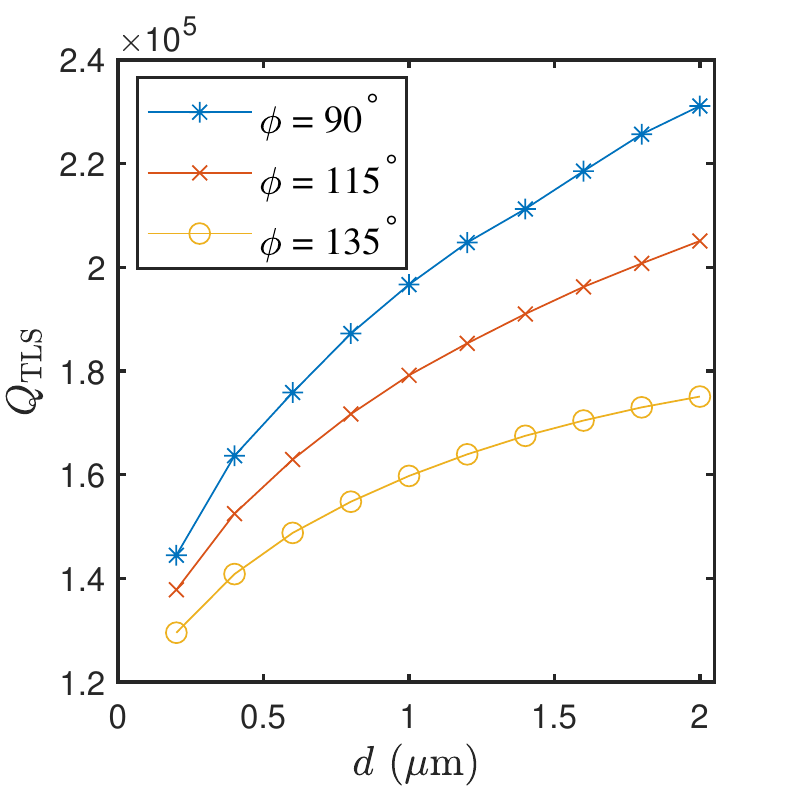}
	\caption{Quality factor $Q_\mathrm{TLS}$ as a function of the trench depth for different indicated sidewall angles. The results are computed from the participation ratios of \fref{fig:p_i}. Here, $w = 10 \textrm{ } \mu\textrm{m}$, $g = 5 \textrm{ } \mu\textrm{m}$, $\mathrm{tan}(\delta_\mathrm{SA}) = \mathrm{tan}(\delta_\mathrm{MS})= \mathrm{tan}(\delta_\mathrm{MA}) = 0.002$, and $\mathrm{tan}(\delta_\mathrm{Si}) = 7.5 \times 10^{-7}$.
	}
	\label{fig:Q_TLS}
\end{figure}

We compute the participation ratios and the resulting $Q_\mathrm{TLS}$ in a CPW structure with the center conductor width $w = 10$~$\mu$m, gap to ground plane $g = 5$~$\mu$m with varying trench depth $d$ and sidewall angle $\phi$ (see \fref{fig:domain} for definitions). Our results are summarized in figures~\ref{fig:p_i} and~\ref{fig:Q_TLS}, and analyzed below.

We observe from \fref{fig:p_i} that the MA interface is the least significant in terms of participation ratios. Whether this interface is significant in terms of losses, depends of course on the loss tangent associated with it. The SA and MS participation ratios are of the order of $10^{-3}$ as opposed to $10^{-5}$ for the MA interface. The  participation ratio of the silicon substrate is the highest, but with the values of loss tangents assumed here, it is not a significant factor in terms of the resulting quality factor. However, if the interfacial loss tangents were one order of magnitude lower than here, the silicon participation would become significant with the assumed loss tangent for the substrate, i.e., $p_i \mathrm{tan} (\delta_i)$ would be within the same order of magnitude in the substrate as in the MS and SA interfaces. 

Figure~\ref{fig:Q_TLS} shows that trenching into the substrate can result in a higher two-level-system-limited quality factor, $Q_\mathrm{TLS}$, by lowering the interfacial and substrate participation ratios significantly. As participation of these regions is decreased, a larger portion of the electric field energy resides in the lossless vacuum, and hence, quality factor is increased. In \fref{fig:p_i} however, the participation ratios show hints of saturation with increasing trench depth, especially for large sidewall angles. Saturation of interface participation with increasing trench depth was observed, e.g., in reference~\cite{Calusine}, as well. Moreover, increasing the sidewall angle increases all shown participation ratios and thus results in higher loss, although for the MA interface this is not yet evident for $\phi$ between 90 to 115 degrees. The effects of increasing the sidewall angle from 90 degrees onward are consistent with the analysis by Wenner~{\it et al.}~\cite{Wenner}. However, they considered only slopes in the sidewalls of the metal conductors and did not consider the effect of the slope on the effectiveness of trenching. Note also that given the definitions of the geometric parameters in \fref{fig:domain}, the value of $\phi$ has an effect on the trench volume, which decreases with increasing $\phi$. This is, however, a distinctively different way of altering the trench volume from, e.g, simply adjusting $g$, due to the resulting differences in the corresponding electric field profiles. Moreover, the sidewall angle has a relatively large effect on the MA participation for small and even vanishing trench depths, because the sidewall angle affects the shapes of the corners of the superconductors located at the ends of the gap. Note that we do not expect the participation ratios and quality factors in \fref{fig:p_i} and \fref{fig:Q_TLS} to necessarily converge to equal values as $d \rightarrow 0$, because even for $d = 0$ the shape of the MA interface is different for different values of $\phi$.

\begin{figure}[!h]
	\centering
	
	\subfloat[][]{\includegraphics[scale=0.675]{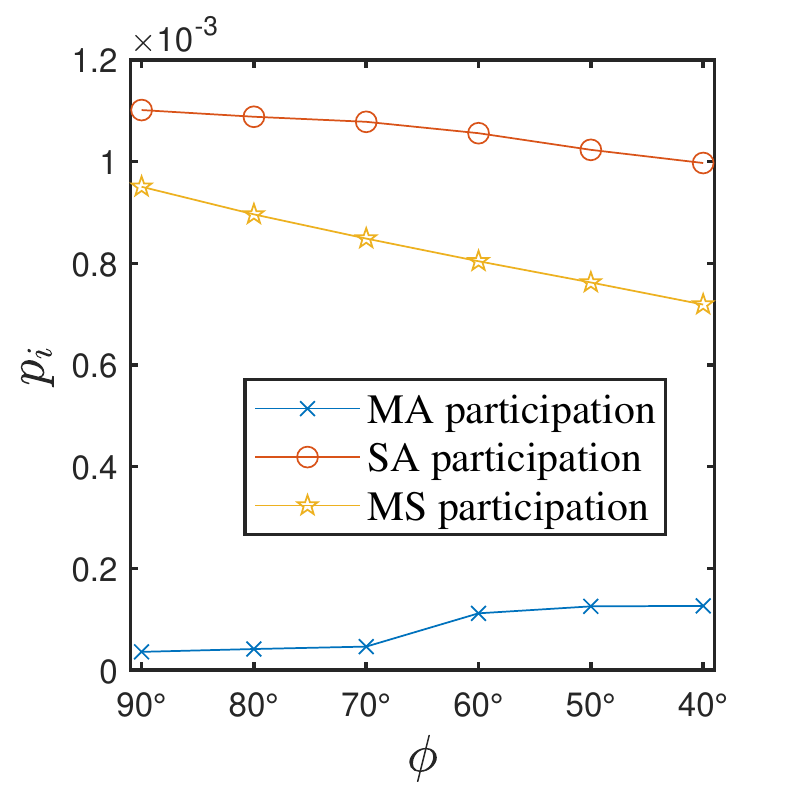}} 
	\subfloat[][]{\includegraphics[scale=0.675]{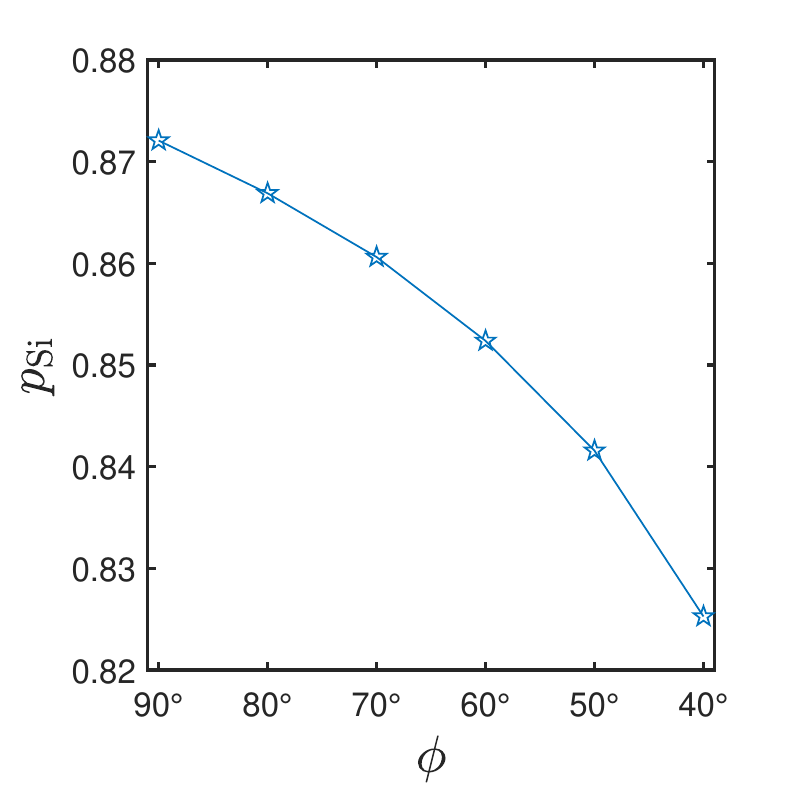}} 
	\subfloat[][\label{fig:Q_TLS-neg-phi}]{\includegraphics[scale=0.675]{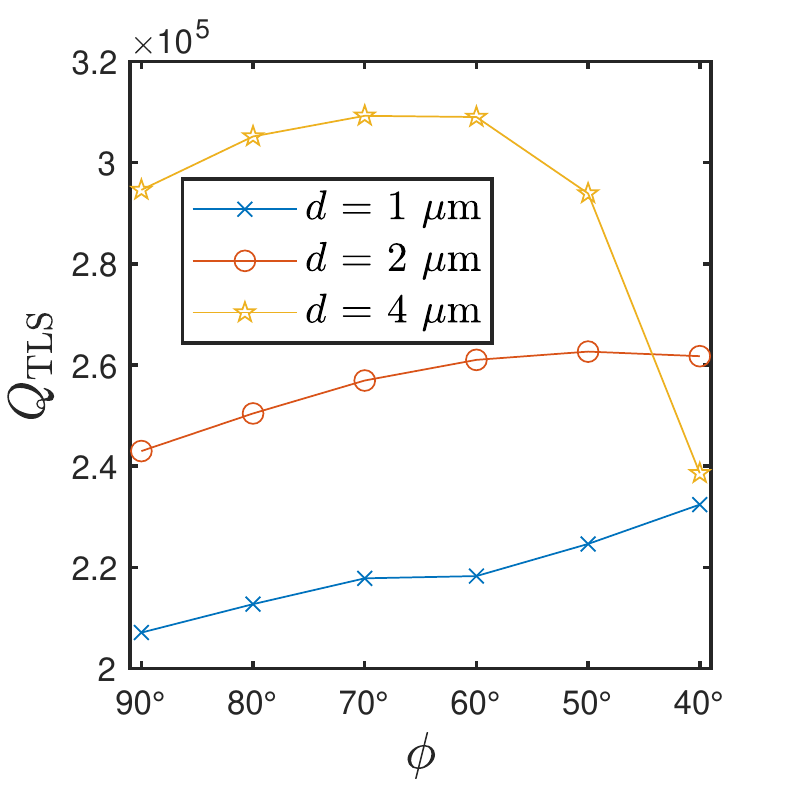}}
	
	\caption{Participation ratios of (a) the  interfaces as indicated and (b) the silicon substrate as functions of decreasing sidewall angle. (c) The resulting quality factor computed from the participation ratios. Here, $d = 1$~$\mu$m (or as indicated), $g = 5$~$\mu$m, $w = 10$~$\mu$m, $\epsilon_\mathrm{SA} = \epsilon_\mathrm{MS} = \epsilon_\mathrm{MA} = 10$, $\epsilon_\mathrm{Si} = 11.6$, $\mathrm{tan}(\delta_\mathrm{SA}) = \mathrm{tan}(\delta_\mathrm{MS}) = \mathrm{tan}(\delta_\mathrm{MA}) = 0.002$, and  $\mathrm{tan}(\delta_\mathrm{Si}) = 7.5 \times 10^{-7}$. }\label{fig:p_i-neg-phi} 
\end{figure}

To investigate also sidewall angles below 90 degrees, we simulate in \fref{fig:p_i-neg-phi} a CPW resonator 
with $\phi$ ranging from 90 to 40 degrees. For this resonator $d = 1$~$\mu$m, $g = 5$~$\mu$m, $w = 10$~$\mu$m, and the height of the superconductors $h = 200$~nm. 
The results indicate that the MS and SA participation ratios, which are the most significant ones of the interfaces in terms of participation, decrease with decreasing sidewall angle. The MA interface exhibits an increasing participation ratio with decreasing angle. This can be attributed to concentration of electric field in the resulting sharp corner. Furthermore, as expected, the substrate participation ratio decreases with decreasing angle, since the smaller the angle, the less we have substrate in the cross-section. Figure~\ref{fig:Q_TLS-neg-phi} shows that the overall effect of decreasing $\phi$ below $90^\circ$ down to $40^\circ$ on $Q_\mathrm{TLS}$ is positive for small trench depths, but not very significant with the used parameters. When the trench depth is moderate, the sidewall is small as well, and hence the effect of the sidewall angle is weak. In figure~\ref{fig:Q_TLS-neg-phi}, we also compute $Q_\mathrm{TLS}$ for sidewall angles of 2 and 4 $\mu$m. For increasing trench depth, the effect of the increasing MA participation becomes more significant, and the optimal angle is found in between 40 and 90 degrees.
 
\begin{figure}[!ht]
	\centering 
	
	\subfloat[][]{\includegraphics[width=7.5cm, height=5.75cm]{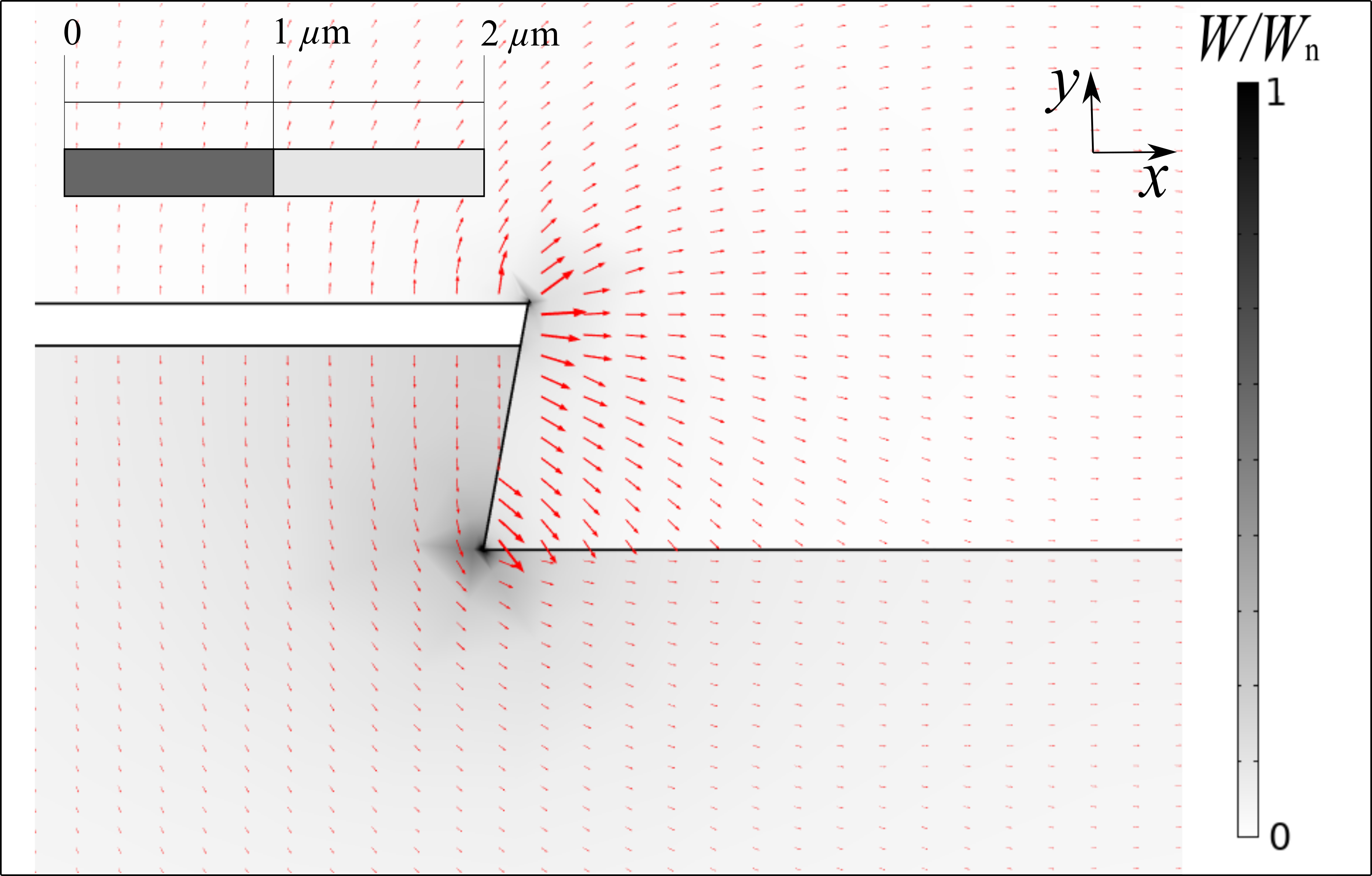}} \hspace{0.075cm}
	\subfloat[][]{\includegraphics[width=7.5cm, height=5.75cm]{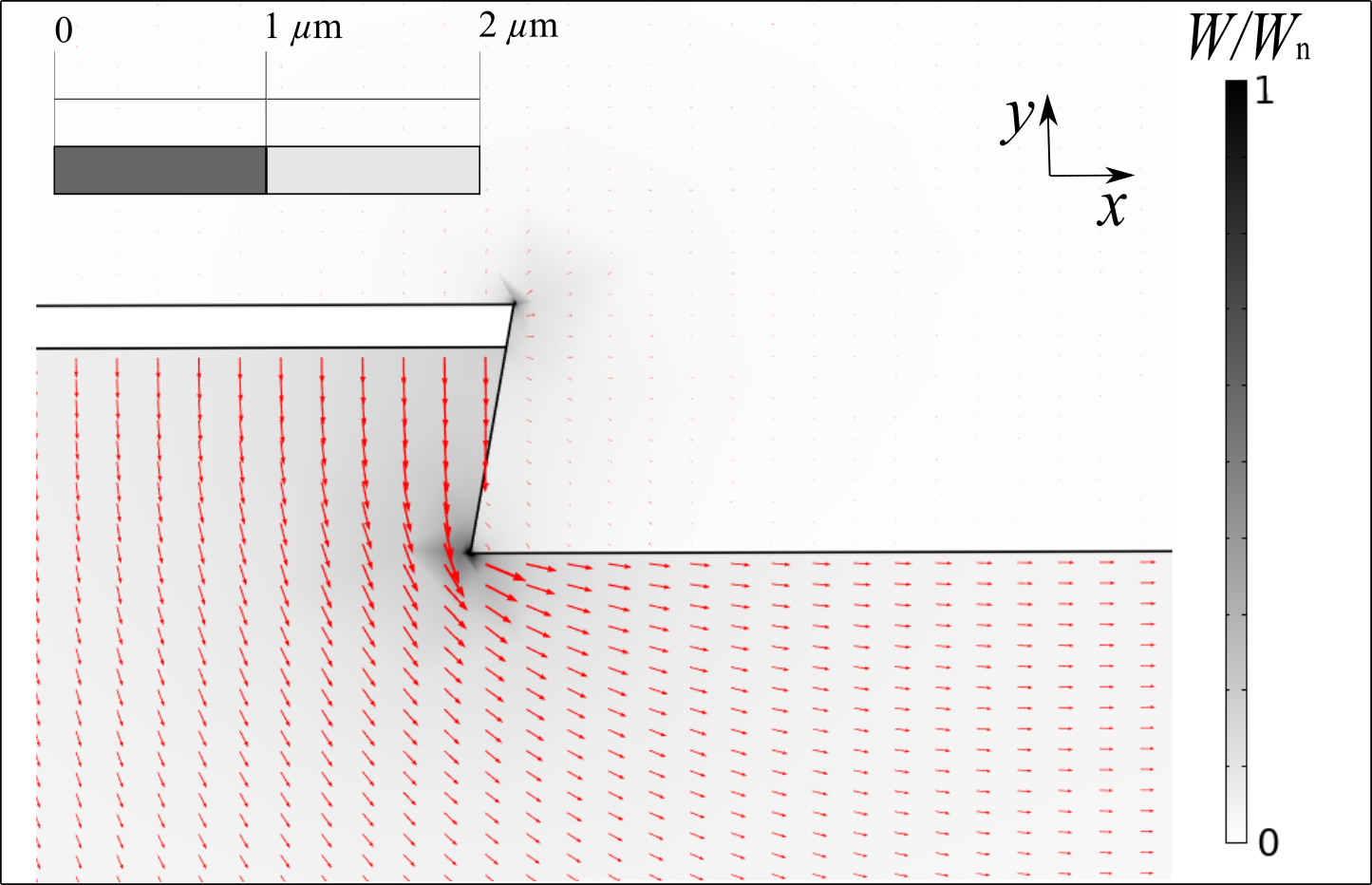}}	
	\caption{(a) Electric field intensity
	${\bf E}$ and (b) electric displacement field ${\bf D}$ ($\bf{D}=\epsilon \bf{E}$) shown by arrows in a part of the resonator cross-section with $\phi = 80 \degree$. The arrows show the direction and relative strength of the fields. The color map represents the electric field energy density $W = \frac{1}{2} \bf{E} \cdot \bf{D}$ normalized to a nominal value $W_\mathrm{n} < \mathrm{max}(W)$ for better visibility of the relative change in $W$. The maxima are located at the dark corners.} \label{fig:fieldProfiles}.
\end{figure}

In \fref{fig:fieldProfiles}, we show the solved electric field for a part of a resonator cross-section. As discussed above, much of the interfacial participation arises from the sharp corners exhibiting high energy densities. Hence, the corners at the material interfaces should be of particular interest in resonator design.

\subsection{Effect of the center conductor width and gap to ground}

In this section, we simulate the effect of the width of the center conductor $w$ and of the gap to ground $g$ together  such that the ratio $w/(w+2g)$ is kept at a constant value of 5/11, resulting in an approximately constant characteristic impedance of roughly $Z_0 = 50~\Omega$ for the CPW \cite{Simons}. In these simulations, the sidewall angle is $\phi=90^\circ$ and we use two different trench depths, $d = 25$~nm and $d = 5$~$\mu$m, the first one representing an almost planar structure, and the second one being an example of a deeply trenched CPW.

\begin{figure}[!h]
	\centering 
	\subfloat[][\label{fig:p_i-d-25nm-a}]{\includegraphics[scale=0.9]{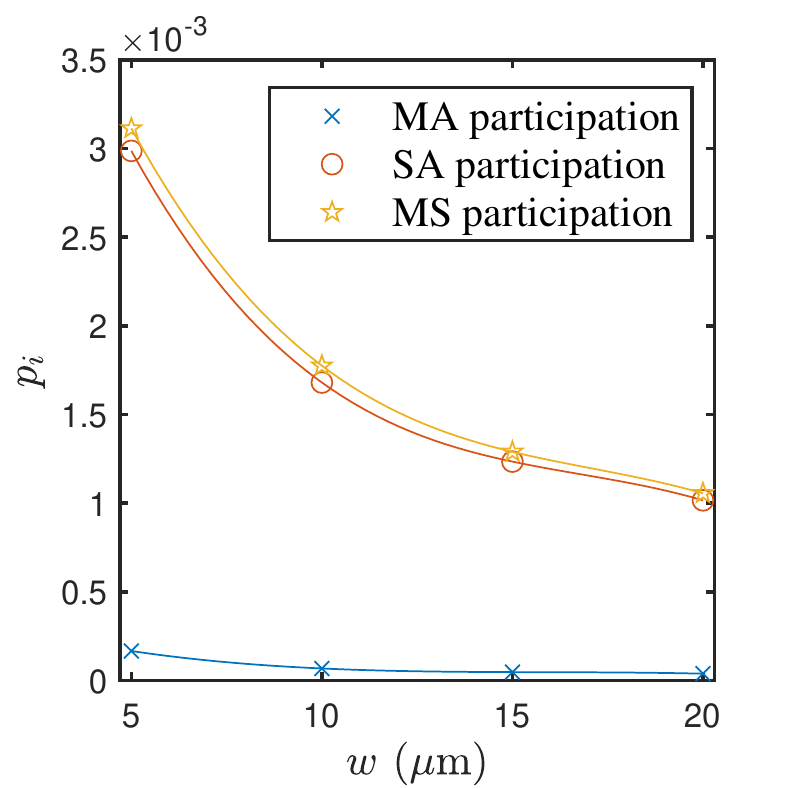}}
	\subfloat[][\label{fig:p_i-d-25nm-b}]{\includegraphics[scale=0.9]{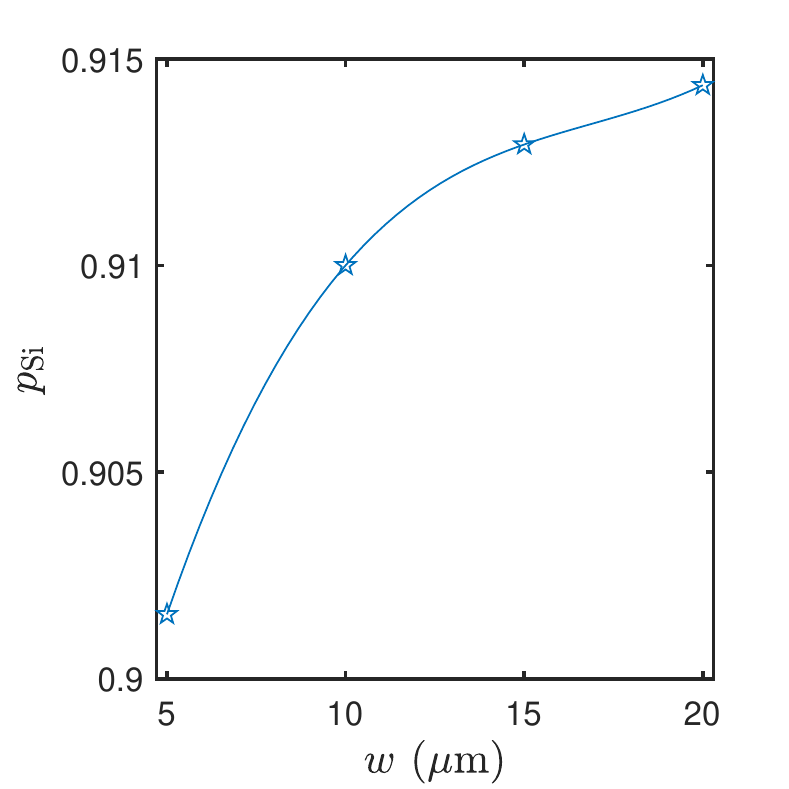}} \\
	\subfloat[][\label{fig:p_i-d-5e-6-a}]{\includegraphics[scale=0.9]{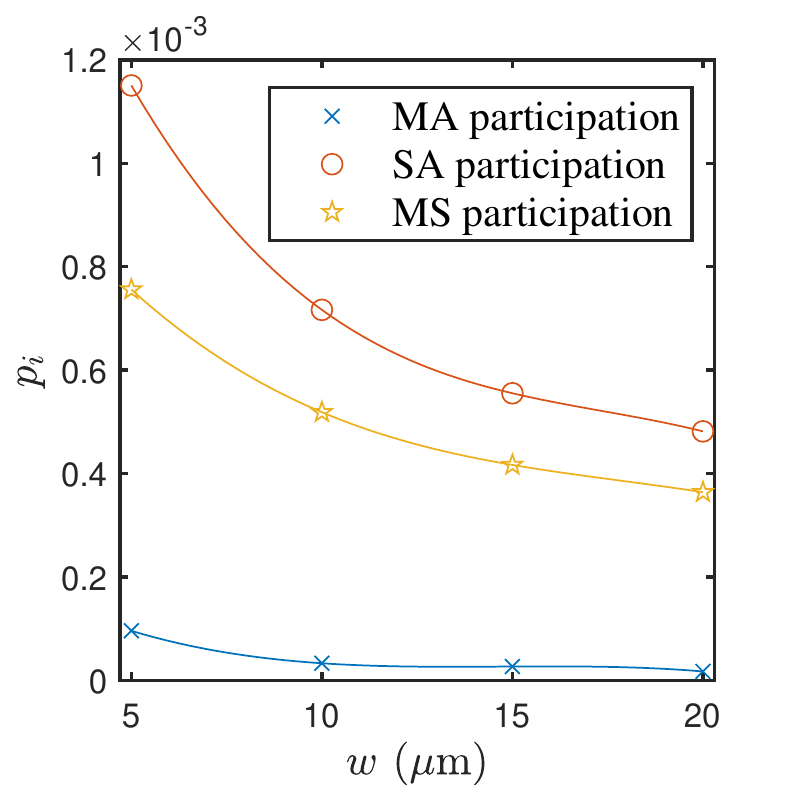}}
	\subfloat[][\label{fig:p_i-d-5e-6-b}]{\includegraphics[scale=0.9]{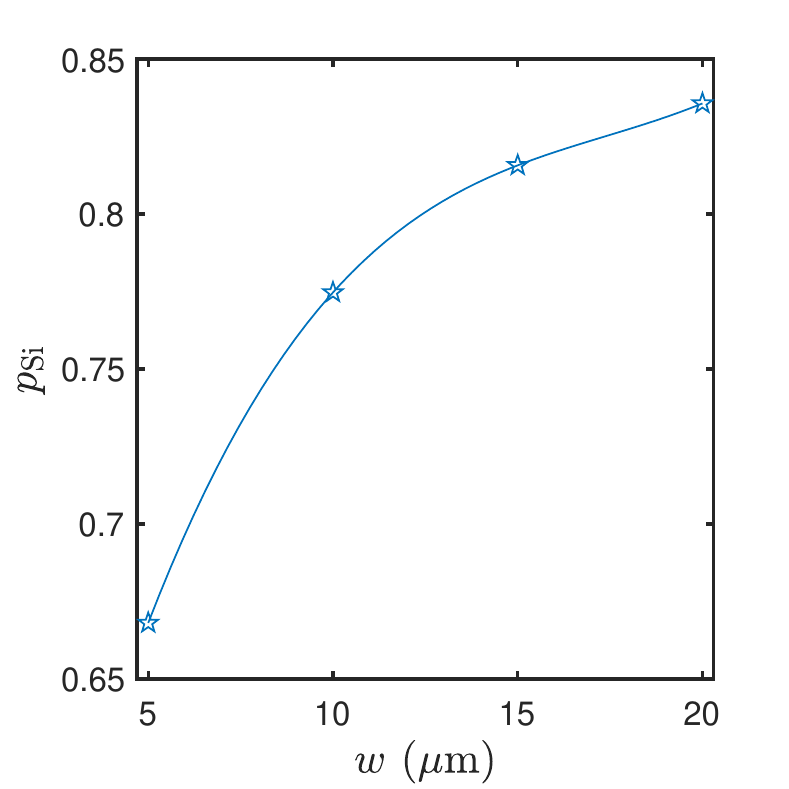}}

	\caption{Participation ratios (markers) (a) and (c) at the interfaces as indicated and (b) and (d) in the substrate as functions of the width of the center conductor $w$. The gap between the center conductor and the ground plane $g$ is adjusted such that $w/(w+2g)=5/11$. The trench depth is $d = 25$~nm in (a) and (b), and $d = 5$~$\mu$m in (c) and (d). The sidewall angle is $90^\circ$ in all panels. The lines between the markers are spline interpolations and merely guides for the eye. $\epsilon_\mathrm{SA} = \epsilon_\mathrm{MS} = \epsilon_\mathrm{MA} = 10$, $\epsilon_\mathrm{Si} = 11.6$.}\label{fig:pi-w-g}
\end{figure}	

\begin{figure}[!h]
	\centering
	
	\subfloat[][\label{fig:Q-d-25nm}]{\includegraphics[scale=0.9]{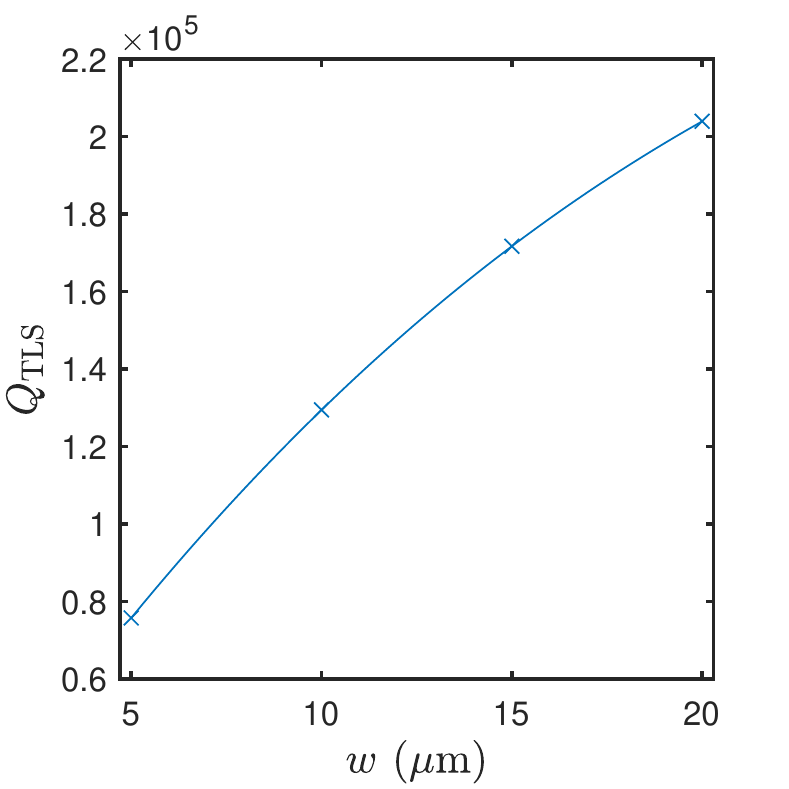}}
	\subfloat[][\label{fig:Q-d-5e-6}]{\includegraphics[scale=0.9]{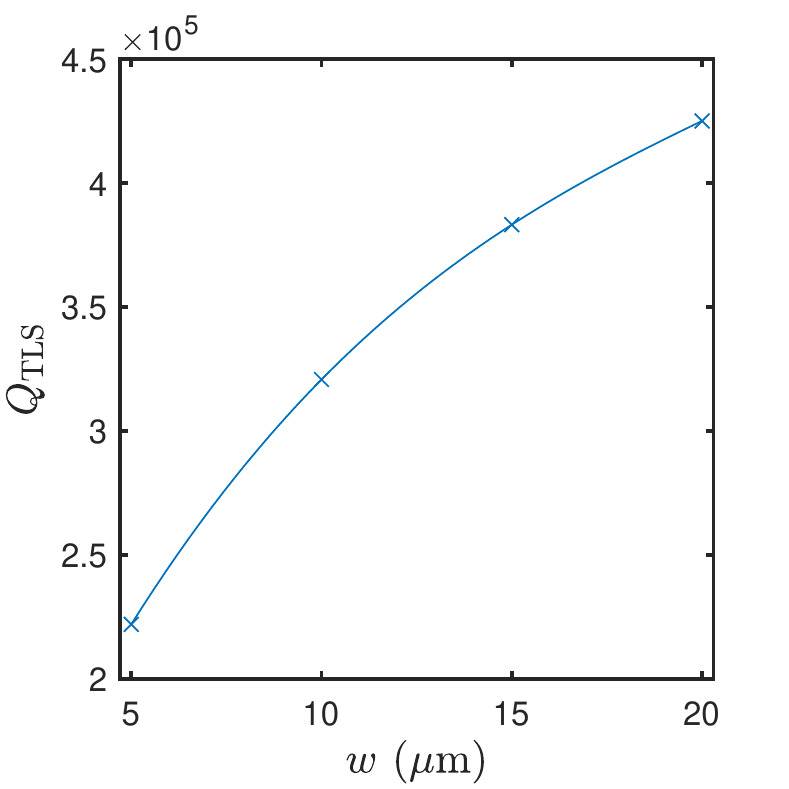}}
	
	\caption{Quality factor $Q_\mathrm{TLS}$ (markers) as a function of the width of the center conductor $w$ such that $w/(w+2g)=5/11$ for (a) $d = 250$~nm and (b) $d = 5$~$\mu$m. The sidewall angle is $90^\circ$. The lines between the markers are spline interpolations and merely guides for the eye. Here,  $\mathrm{tan}(\delta_\mathrm{SA}) = \mathrm{tan}(\delta_\mathrm{MS}) = \mathrm{tan}(\delta_\mathrm{MA}) = 0.002$, and $\mathrm{tan}(\delta_\mathrm{Si}) = 7.5 \times 10^{-7}$.}\label{fig:Q_TLS-w-g}
\end{figure}

From \fref{fig:pi-w-g}, we observe that widening the 50 $\Omega$ CPW structure significantly reduces participation at the interfaces: The SA and MS participation ratios are almost halved by doubling the width of the structure from  $w = 5$~$\mu$m to $w = 10$~$\mu$m. For even further widening, this effect becomes less drastic but remains significant. On the other hand, the substrate participation increases with increasing wideness: the increase in $p_\mathrm{Si}$ from $w = 5$~$\mu$m to $w = 20$~$\mu$m when $d = 25$~nm is 0.013. 
Thus the participation ratio of the lossless vacuum is decreasing 
which is not in general desired.
Fortunately, with the assumed loss tangents, the resulting total effect on $Q_\mathrm{TLS}$ is positive, as shown \fref{fig:Q_TLS-w-g}. Again, we observe that the deeper the trench, the lower the dielectric loss. 

\pagebreak

\subsection{Effect of the dielectric constants}

Next, we turn our attention to the effect of the dielectric constants of the lossy interface regions on the resulting dielectric losses. Given participation ratios $p^{\mathrm{sim}}_i$ computed as results of a finite-element simulation with given dielectric constants of the interface regions $\epsilon^{\mathrm{nom}}_\mathrm{i}$, the following assumption has been made in previous studies~\cite{Calusine,Woods}:  assuming constant thicknesses for the lossy interface regions and a fixed dielectric constant for the substrate, losses can be essentially scaled by scaling the participation ratios with the dielectric constants as\footnote{In references~\cite{Calusine, Woods} this  scaling of the losses was essentially embedded in the concept of \emph{loss factor}, combining loss tangents and the scaling of participation ratios.}

\begin{equation}\label{SAScale}
p_\mathrm{SA} = \frac{\epsilon_\mathrm{SA}}{\epsilon^{\mathrm{nom}}_\mathrm{SA}} p^{\mathrm{sim}}_\mathrm{SA},
\end{equation}
\begin{equation}\label{MSScale}
p_\mathrm{MS} = \frac{\epsilon^{\mathrm{nom}}_\mathrm{MS}}{\epsilon_\mathrm{MS}} p^{\mathrm{sim}}_\mathrm{MS},
\end{equation}
\begin{equation}\label{MAScale}
p_\mathrm{MA} = \frac{\epsilon^{\mathrm{nom}}_\mathrm{MA}}{\epsilon_\mathrm{MA}} p^{\mathrm{sim}}_\mathrm{MA}.
\end{equation}
This assumption offers the opportunity to use the loss resulting from a simulation obtained using $ \epsilon^{\mathrm{nom}}_\mathrm{i}$ to estimate the loss for any other value of $\epsilon_i$ without additional simulations. For this assumption to hold, however, the loss contribution in one region has to be unaffected by the changes in loss contributions of the other regions. This is an approximation and does not hold exactly, but let us investigate the validity of this assumption in practice.

To study how variations in the dielectric constants affects the dielectric losses, and to study the validity of equations~\eqref{SAScale}--\eqref{MAScale}, we simulate participation ratios for a CPW resonator with the sidewall angle $\phi$ varying from 90$^\circ$ to 140$^\circ$  for different combinations of the interfacial dielectric constants using $\epsilon_\mathrm{Si} = 11.6$ for the dielectric constant of the substrate. In these simulations, $d = 250$~nm, $g = 6$~$\mu$m, and $w = 10$~$\mu$m. We assume the thickness of each lossy dielectric interface to be 5~nm. We fix $\epsilon_\mathrm{SA} = 10$ and vary $\epsilon_\mathrm{MS}$ and  $\epsilon_\mathrm{MA}$ between values 5, 10, and 15.

\begin{figure}[!h]
	\centering 
	
	\subfloat[][]{\includegraphics[scale=0.9]{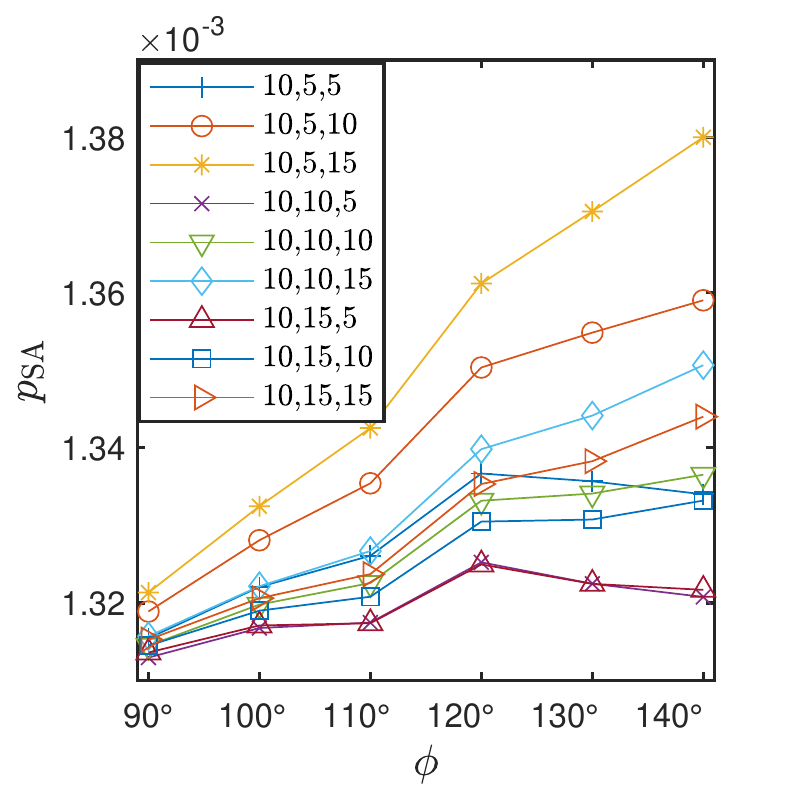}}
	\subfloat[][]{\includegraphics[scale=0.9]{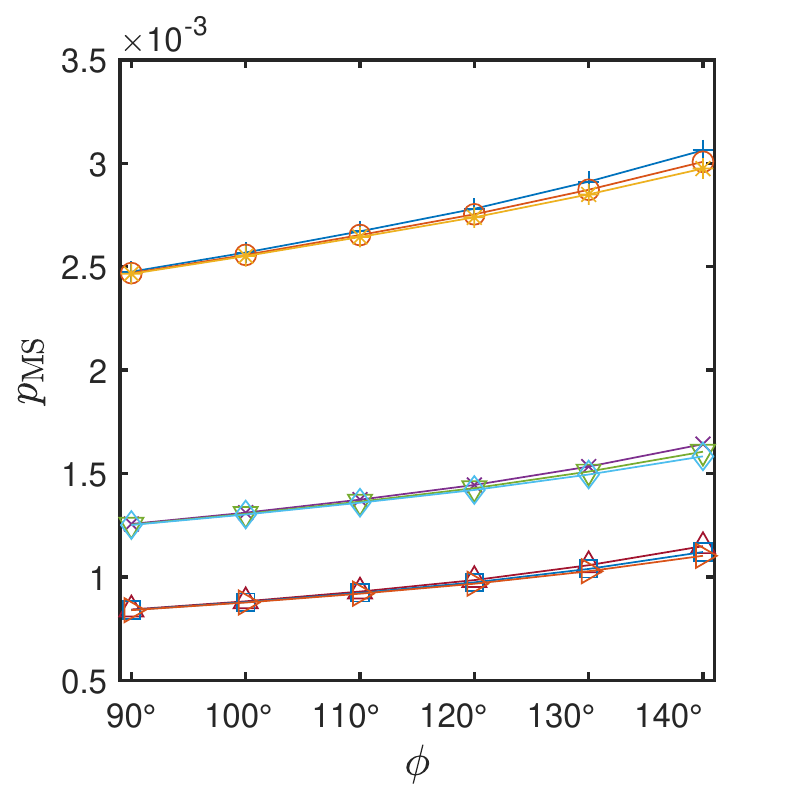}} \\
	\subfloat[][]{\includegraphics[scale=0.9]{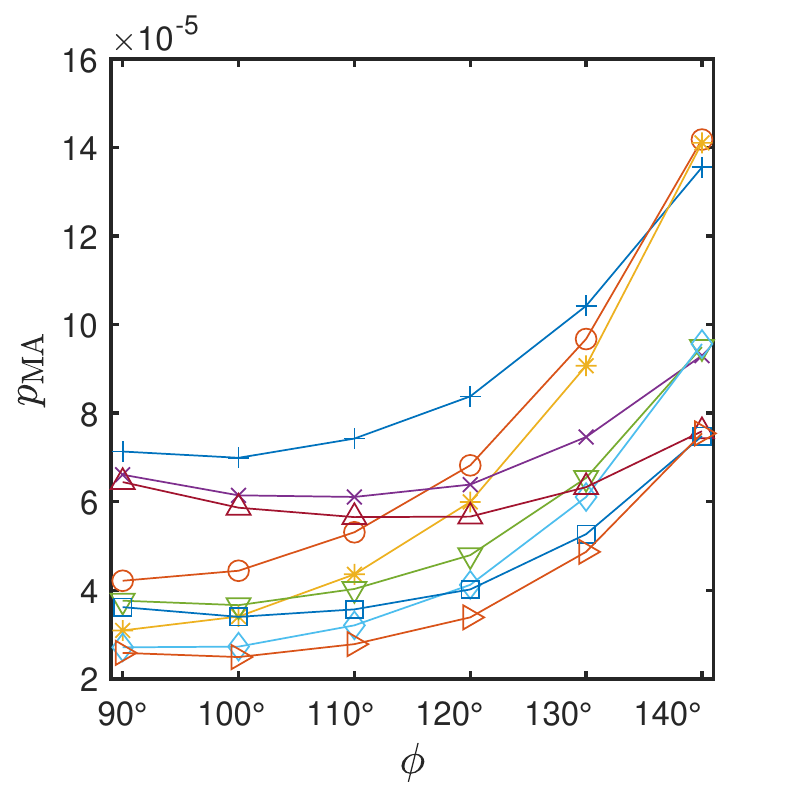}}
	\subfloat[][]{\includegraphics[scale=0.9]{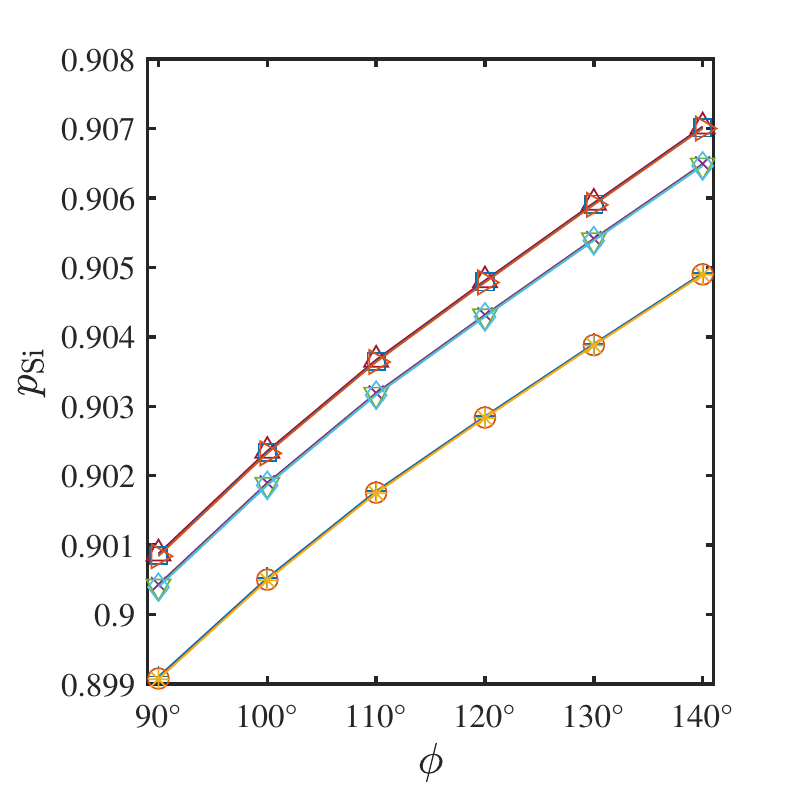}}
	
	\caption{Participation ratios of (a) substrate-to-air, (b) metal-to-substrate, and (c) metal-to-air interfaces and (d) of the substrate for different dielectric constants  as functions of the sidewall angle. 
	The legend denotes the dielectric constants in the following order: $\epsilon_\mathrm{SA}$,  $\epsilon_\mathrm{MS}$,  $\epsilon_\mathrm{MA}$. Note that in (b) and in (d) the curves with equal $\epsilon_\mathrm{MS}$ overlap. Geometric parameters: $d = 250$~nm, $g = 6$~$\mu$m, and $w = 10$~$\mu$m.}\label{fig:p_i-eps}
\end{figure}	
The simulated participation ratios are shown in \fref{fig:p_i-eps}. We observe that relative to the value of the participation ratio, those for SA, MS, and MA interfaces depend on the dielectric constants of the other interfaces. The change is less than 10\% for SA and MS participation. For MA interface with $\epsilon_\mathrm{MA} = 15$, $\epsilon_\mathrm{SA} = 10$, however, the case of $\epsilon_\mathrm{MS} = 5$ results in 87\% higher participation than the case of  $\epsilon_\mathrm{MS} = 15$ at sidewall angle $\phi = 140 \degree$.
Note that relative changes in the substrate participation arising from the changes in the interfacial dielectric constants are negligible, which is expected since the substrate participation is orders of magnitude larger than the interface participation. 

The results in \fref{fig:p_i-eps} also indicate, that equations \eqref{SAScale}--\eqref{MAScale} do not necessarily yield the correct result for the participation ratios with changing dielectric constants. To study this discrepancy in more detail, we computed the participation ratios $p_i$ for a changing $\epsilon_{i}$ while keeping the other $\epsilon_j$ constant. The results are compared with those obtainable from equations~\eqref{SAScale}--\eqref{MAScale} in table~\ref{table:scalingResults}. It seems that equation~\eqref{MSScale} predicts the scaling of $p_\mathrm{MS}$ with $\epsilon_\mathrm{MS}$ very well. The scaling with $\epsilon_\mathrm{SA}$ shows slightly more deviation from equation~\eqref{SAScale}, which nonetheless seems to be a decent approximation. However, equation~\eqref{MAScale} exhibits poor behavior as the sidewall angle $\phi$ is increased---it appears that equation~\eqref{MAScale} cannot be used as a reliable approximation at all for $\phi = 140 \degree$.

\begin{table}[h!]
	\begin{center} 
		\caption{Interface participation ratios for different dielectric constants. The values of the changing $\epsilon_i$ are indicated, whereas the other dielectric constants equal $10$. The results in each cell are normalized to the value in the square brackets. The values in parentheses are obtained using equations \eqref{SAScale}--\eqref{MAScale}.}
		\label{table:scalingResults} \small
		\begin{tabular}{|c|c|c|} 
			
			\multicolumn{1}{c}{} & \multicolumn{1}{c}{$\phi = 90 \degree$} & \multicolumn{1}{c}{$\phi = 140 \degree$} \\
			\hline
			\text{$\epsilon_\mathrm{SA}$ } & \text{$p_\mathrm{SA}$} \text{ [7.007 $\times$ $10^{-4}$]} & \text{$p_\mathrm{SA}$} \text{ [6.882 $\times$ $10^{-4}$]}  \\
			\hline
			5 & 1.00 \text{(1.00)} &  1.00 \text{(1.00)} \\
			10 & 1.88 \text{(2.00)} & 1.94 \text{(2.00)} \\
			15 & 2.75 \text{(3.00)} & 2.87 \text{(3.00)} \\
			\hline
			\text{$\epsilon_\mathrm{MS}$} & \text{$p_\mathrm{MS}$} \text{ [2.469 $\times$ $10^{-3}$]}& \text{$p_\mathrm{MS}$} \text{ [3.007 $\times 10^{-3}$]} \\
			\hline
			5 & 1.00 (1.00)  & 1.00 (1.00)  \\
			10 & 0.51 \text{(0.50)} & 0.53 \text{(0.50)} \\
			15 & 0.34 \text{(0.33)}  & 0.37 \text{(0.33)} \\
			\hline
			\text{$\epsilon_\mathrm{MA}$}  & \text{$p_\mathrm{MA}$} \text{ [6.610 $\times$ $10^{-5}$]} & \text{$p_\mathrm{MA}$} \text{ [9.302 $\times$ $10^{-5}$]} \\
			\hline
			5 & 1.00 \text{(1.00)} & 1.00 \text{(1.00)}  \\
			10 & 0.57 \text{(0.50)} & 1.02 \text{(0.50)} \\
			15 & 0.41 \text{(0.33)} & 1.03 \text{(0.33)} \\
			\hline
		\end{tabular}
	\end{center}
\end{table} 

\subsection{Concluding remarks}

Trenching into the substrate is an effective way to decrease the interfacial and bulk substrate participation ratios in CPW structures. Furthermore, widening the CPW structure while keeping the desired characteristic impedance can reduce losses. Sloped sidewalls with a sidewall angle $\phi < 90 \degree$ seem to decrease the dielectric loss. Sloped sidewalls with $\phi > 90 \degree$ in the trenched gap, however, not only increase losses, but seem to decrease the effectiveness of trenching. It is thus advantageous to control the slopes of the sidewalls in the fabrication phase. However, if the MA interface material is inherently much more lossy than the rest of the lossy regions, as seems to be the case in the devices analyzed in reference~\cite{Calusine}, fabricating sidewall angles below 90 degrees may not be very effective, since the MA participation increases with decreasing sidewall angle. Assuming equal thicknesses, loss tangents, and dielectric constants for SA, MS, and MA interfaces, the MA interface is not a significant source of loss: most of the interface loss comes from the SA and MS interfaces. 

The dielectric constants affect the participation ratios of the lossy regions and thus the dielectric losses. The participation ratio of a given region is not independent of the dielectric constants of the other regions either. While simply scaling the losses of the SA and MS interfaces using equations \eqref{SAScale} and \eqref{MSScale} as the dielectric constant of one interface region is varied provides a good approximation, neglecting variation in the dielectric constants of the other regions can introduce further errors. The MA participation, on the other hand, is particularly sensitive to changes in the dielectric constants of the SA and MS interfaces. Furthermore, the predictive capability of equations \eqref{SAScale}--\eqref{MAScale} seems to depend on the sidewall angle of the CPW: the worst approximation is provided by equation~$\eqref{MAScale}$ for large angles, likely due to the resulting sharp corners in the superconductor. A partial remedy for this could be treating the resulting corners separately. 

Our observations in this paper thus far serve as a motivation for accurately finding the dielectric constants for the different regions, such that the finite-element models reproduce given measurement data. This is the topic of the following section.

\section{Method to obtain dielectric constants and loss tangents}

In this section, we present a method for solving for the dielectric constants and loss tangents of the different regions of CPW resonator samples utilizing typically accessible measurement data. This method serves as a tool for inferring these material parameters of the devices, and as motivated in section~3, this information is necessary for accurate estimation of the loss. We assume that the material properties are equal for a set of different samples. Whether this is a justified assumption, depends of course, for example, on the fabrication process. However, samples fabricated in the same batch or even on the same chip, have typically reasonably small variations in their material properties. Our techniques are similar to those utilized, e.g., in references~\cite{Calusine, Woods}.

\subsection{Finding dielectric constants from resonance frequency measurements}

Ignoring coupling with transmission line, the fundamental resonance frequency of a half-wave-length CPW resonator is given by
\begin{equation}\label{f0}
f_0 = \frac{c}{\sqrt{\epsilon_\mathrm{eff}} 2 l},
\end{equation}
where $c$ is the speed of light in vacuum, 
$l$ is the length of the resonator, and $\epsilon_\mathrm{eff}$ is the effective dielectric constant. While analytical expressions for $\epsilon_\mathrm{eff}$ exist in simple geometries \cite{Simons}, we find it from our finite-element simulations using the definition
\begin{equation}\label{eps_eff}
\frac{1}{2}\epsilon_\mathrm{eff} \int_\Omega \epsilon_0 \| {\bf E} \|^2 \rmd A = \frac{1}{2} \int_\Omega \epsilon \epsilon_0 \| {\bf E} \|^2 \rmd A, 
\end{equation}
where $\epsilon_0$ is the permittivity of vacuum. Thus $\epsilon_\mathrm{eff}$ is the dielectric constant of a hypothetical homogeneous material filling the whole cross-section of the resonator and resulting in an electric field energy equal to that of the actual spatially distributed dielectric constant $\epsilon$, under quasi-static approximation. With our assumptions of linear materials
we may rewrite equation~\eqref{eps_eff} as
 \begin{equation}\label{eps_eff2}
\epsilon_\mathrm{eff} = \frac{\sum^n_{i=1}\int_{\Omega_i} \epsilon_i \| {\bf E} \|^2 \rmd A}{\int_\Omega \| {\bf E} \|^2 \rmd A} = \frac{\sum^n_{i=1} \epsilon_i \int_{\Omega_i} \| {\bf E} \|^2 \rmd A}{\int_\Omega \| {\bf E} \|^2 \rmd A} =: \sum^n_{i=1} \epsilon_i F_i,
\end{equation}
where we refer to $F_i$ as the \emph{filling factor}.\footnote{The terms \emph{filling factor} and \emph{participation ratio} are often used somewhat interchangeably in the literature. Here, however, we make this distinction to differentiate between $p_i$ and $F_i$.} The definition of the filling factor resembles that of the participation ratio, but there is a distinctive difference on how the permittivity appears inside the integral. 

Equations \eqref{f0} and \eqref{eps_eff2} yield a connection between the dielectric constants of the different regions $\Omega_i$ and the resonance frequency of the resonator:
\begin{equation}\label{eps_eff-constraint}
\sum^n_{i=1} \epsilon_i F_i = \frac{c^2}{f^2_0 4 l^2}. 	
\end{equation}
Assuming the dielectric constants of the SA, MS, and MA interfaces and of the substrate were unknown but equal from device to device, we can potentially solve for $\epsilon_i$ given resonance frequency measurements for four CPW resonator samples, by solving the nonlinear matrix equation
\begin{equation}\label{eps_eff-constraint-matrix}
\mathrm{F} \pmb{\epsilon} = \pmb{ \epsilon}_\mathrm{\bf eff}, 	
\end{equation}
where each row of $\mathrm{F}$, which depends on $\pmb{\epsilon}$, represents the filling factors for all regions $\Omega_i$ for the sample corresponding to the row, obtained from field solutions, $\pmb{\epsilon}$ is the vector of dielectric constants, with the dielectric constant of vacuum $\epsilon_\mathrm{vac} = 1$ as one of the components, and the components of the vector $\pmb{ \epsilon}_\mathrm{\bf eff}$ are $\epsilon^i_\mathrm{eff} =\frac{c^2}{{f_0^i}^2 4 l^2}$, and they are obtained from measured resonance frequencies $f^i_0$ of the samples. Thus in the case of four samples and five regions with distinct dielectric constants, $\mathrm{F}$ is a 4-by-5 matrix, $\pmb{\epsilon}$ is a 5-by-1 vector and  $\pmb{ \epsilon}_\mathrm{\bf eff}$ is a 4-by-1 vector. Since $\epsilon_\mathrm{vac} = 1$ is known, this results in an equal number of unknowns as equations. 

\subsection{Finding loss tangents from quality factor measurements}

In brief, given measurement data of quality factors for $n-1$ CPW resonators, one can find the loss tangents of $n-1$ lossy regions, in addition to the lossless vacuum, in a similar manner as solving for the dielectric constants. Assuming equal loss tangents for the different material regions from device to device, we can solve the matrix form of equation~\eqref{Q_TLS}
\begin{equation}\label{Q_TLS-matrix}
\mathrm{P} \mathrm{tan}(\pmb{\delta})= \frac{1}{\pmb{Q}_\mathrm{\bf TLS}},
\end{equation}
where $\mathrm{P}$ is the participation matrix obtained from field solutions, each of its rows representing the participation ratios for all regions $\Omega_i$ for the corresponding sample, $\mathrm{tan}(\pmb{\delta})$ is the vector of loss tangents, and the components of the vector $\frac{1}{\pmb{Q}_\mathrm{\bf TLS}}$ are obtained from the quality factor measurements for each sample. Thus, in the case of four samples and four lossy dielectric regions, the size of $\mathrm{P}$ is 4-by-4, and both   $\mathrm{tan}(\pmb{\delta})$ and $\frac{1}{\pmb{Q}_\mathrm{\bf TLS}}$ are 4-by-1 vectors. In this work, we define all functional operations on vectors componentwise.

\subsection{Solving the inverse problem: series and parallel approaches}

As discussed above, given measurement data for $f_0$ and $Q_\mathrm{TLS}$ for a set of CPW resonators, it is possible find the dielectric constants and loss tangents which reproduce such data from simulations. However, the matrices $\mathrm{P}$ and $\mathrm{F}$ depend on the field solutions and thus on the dielectric constants of the lossy regions. Hence, as the dielectric constants are generally unknown, solving for these values is not simply a matter of inverting $\mathrm{P}$ and $\mathrm{F}$; we do not know their elements in advance.

Consequently, we formulate the inverse problem of finding $\mathrm{tan}(\pmb{\delta})$ and $\pmb{\epsilon}$ for a set of CPW resonators, each of them having $n-1$ lossy dielectric regions, in two ways. We call these the \emph{series} and \emph{parallel} approaches to this problem.

\begin{myproblem}[Series approach] Let $\pmb{ \epsilon}_\mathrm{\bf eff}$ and $\pmb{Q}_\mathrm{\bf TLS}$ be vectors of effective dielectric constants and quality factors for a set of CPW resonators, respectively, obtained as measurement data. Let $\pmb{ \epsilon}^\mathrm{\bf sim}_\mathrm{\bf eff}$ and $\pmb{Q}^\mathrm{\bf sim}_\mathrm{\bf TLS}$ be the corresponding vectors computed from the field solutions using equations~\eqref{eps_eff-constraint-matrix} and \eqref{Q_TLS-matrix}.
	
\begin{enumerate}
	\item Find $\pmb{\epsilon}$ that minimizes $\| \pmb{\epsilon}_\mathrm{\bf eff} - \pmb{\epsilon}^\mathrm{\bf sim}_\mathrm{\bf eff} \|^2$, such that $\epsilon^\mathrm{L}_i \leq \epsilon_i \leq {{\epsilon^\mathrm{U}_i}}$ for all $i \in \{1,2,..,,n-1\}$.
	\item Using the found $\pmb{\epsilon}$, find $\mathrm{tan}(\pmb{\delta})$ that minimizes $\|\pmb{Q}_\mathrm{\bf TLS} - \pmb{Q}^\mathrm{\bf sim}_\mathrm{\bf TLS} \|^2$, such that ${{\mathrm{tan}(\delta_i)^\mathrm{L}}} \leq \mathrm{tan}(\delta_i) \leq {{\mathrm{tan}(\delta_i)^\mathrm{U}}}$ for all $i \in \{1,2,...,n-1\}$.
\end{enumerate}	

\noindent Here, ${{\epsilon^\mathrm{L}_i}}$, ${{\epsilon^\mathrm{U}_i}}$, ${{\mathrm{tan}(\delta_i)^\mathrm{L}}}$ and ${{\mathrm{tan}(\delta_i)^\mathrm{U}}}$ denote the lower and upper bounds for each dielectric constant and loss tangent in the optimization, respectively.
\end{myproblem}

\noindent In this approach, part (i) is the time-consuming one, since one needs to solve the field problems at every objective function evaluation. Once part (i) has been solved, the objective function in part (ii) is much less computationally expensive, and thus the second step of the series approach is relatively fast, as only a single set of field solutions is required. Note that within the quasistatic approach employed here, the loss tangents do not affect the resonance frequencies and hence part (i) does not need input from part (ii). In experiments, the small losses do not either have a strong effect on the resonance frequencies.

However, in essence we are dealing with a multi-objective optimization problem: we want to minimize the error with respect to quality factor measurements and resonance frequency measurements. Instead of minimizing the objectives in series, one can reformulate such a multi-objective optimization problem as a single-objective optimization problem by, e.g., minimizing the weighted sum of the two objectives. Another option is to reformulate the multi-objective optimization problem as a constrained single-objective problem \cite{Emmerich}. To avoid the question of choosing the weight for each objective, we take the latter approach and minimize in terms of quality factors but restrict the resonance frequencies as additional constraints in the optimization problem.

\begin{myproblem}[Parallel approach] Let $\pmb{\epsilon}_\mathrm{\bf eff}$ and $\pmb{Q}_\mathrm{\bf TLS}$ be vectors obtained as measurement data. Let $\pmb{Q}^\mathrm{\bf sim}_\mathrm{\bf TLS}$ be computed from the field solution using equation~\eqref{Q_TLS-matrix}.
\begin{itemize}	
	\item Find $\pmb{\epsilon}$ and $\mathrm{tan}(\pmb{\delta})$ that minimize $\| \pmb{Q}_\mathrm{\bf TLS} - \pmb{Q}^\mathrm{\bf sim}_\mathrm{\bf TLS} \|^2$ subject to the nonlinear constraints $\vert \left(\mathrm{F} \pmb{\epsilon}\right)_i - \epsilon_\mathrm{eff}^i \vert \leq \tau$, such that ${{\epsilon^\mathrm{L}_i}} \leq \epsilon_i \leq {{\epsilon^\mathrm{U}_i}}$ and ${{\mathrm{tan}(\delta_i)^\mathrm{L}}} \leq \mathrm{tan}(\delta_i) \leq {{\mathrm{tan}(\delta_i)^\mathrm{U}}}$ for all $i \in \{1,2,...,n-1\}$,
\end{itemize}	
where $\tau$ is a chosen tolerance for the constraints.
\end{myproblem}

\noindent Hence, in the parallel approach, the vector of optimization variables consists of the components of $\pmb{\epsilon}$ and $\mathrm{tan}(\pmb{\delta})$. 

This approach has the advantage of simultaneously requiring the dielectric constants to satisfy the quality factor data and the resonance frequency data, as they affect both these values through simulations. The disadvantage is that the number of optimization variables is increased compared to the individual steps of the series approach. Here not only the objective function evaluations require solving the field problem for the whole set of samples at every iteration, but also the evaluation of the constraint needs this information. Fortunately, since the objective function evaluation yields the field solutions for a given $\pmb{\epsilon}$, they can be reused whenever the solutions corresponding to an indentical $\pmb{\epsilon}$ are required.

\section{Case study} 

Above, we presented the principle for solving the dielectric constants and loss tangents of the different lossy regions utilizing a set of CPW resonators. However, in practice, the matrices $\mathrm{P}$ and $\mathrm{F}$ can have very high condition numbers of the order of $10^5$ to $10^6$ if attention is not paid to this issue. This is the case, for example, for rather typical sets of anisotropically etched samples: the relative changes in the interface participation ratios are almost equal for changes in the device geometry, producing nearly linearly dependent rows for the participation matrix \cite{Calusine, Woods}. This behavior appears also for the filling factors which essentially constitute a special case of participation ratios. 

In practice, a high condition number $\kappa\left(\mathrm{P}\right) = \|\mathrm{P}^{-1}\| \|\mathrm{P}\|$ translates to outputs of the model being highly sensitive to the inputs. In terms of equation~\eqref{Q_TLS-matrix}, this means that even relatively small noise in the measurement data of $Q_\mathrm{TLS}$, or small changes in $\mathrm{P}$, can cause high uncertainty in the resulting loss tangents. In reference~\cite{Woods}, this problem was addressed by forming a set of CPW resonators that produce a much lower condition number of the order of $10^3$ by utilizing isotropic etching.

In this section, we present a case study and solve the above-formulated inverse problems for a set of CPW resonators. We design a set of cross-sections that produces $\mathrm{P}$ and $\mathrm{F}$ with relatively low condition numbers. 

\subsection{The modeled cross-sections}

\begin{figure}[!t]
	\centering
	\subfloat[]{\includegraphics[scale=0.33]{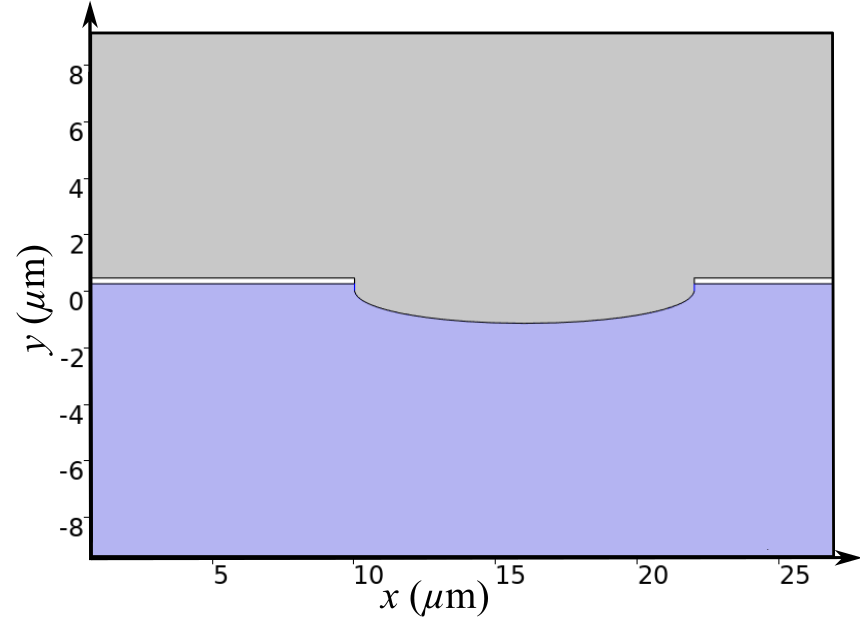}}\hspace{0.0003cm}
				\subfloat[]{\includegraphics[scale=0.33]{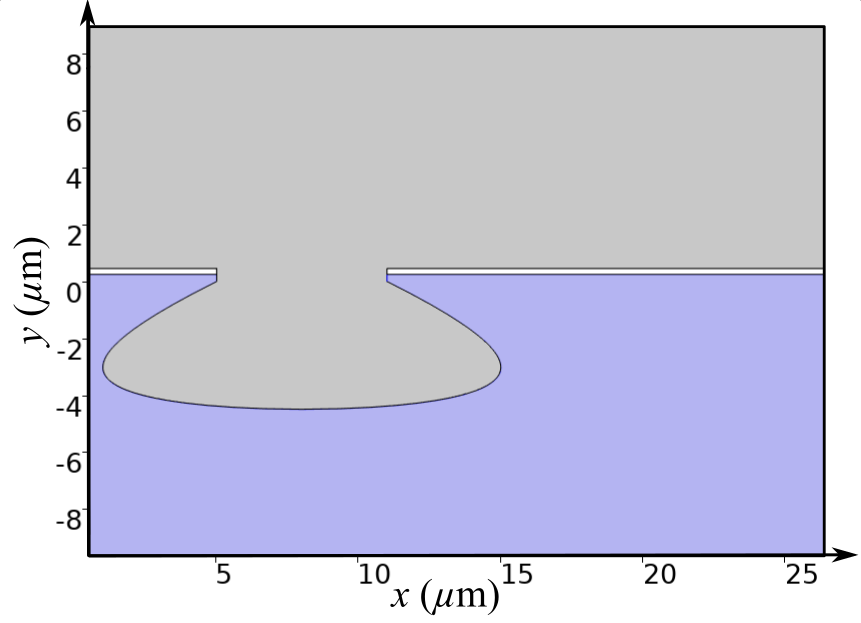}}  	\\
	\subfloat[]{\includegraphics[scale=0.33]{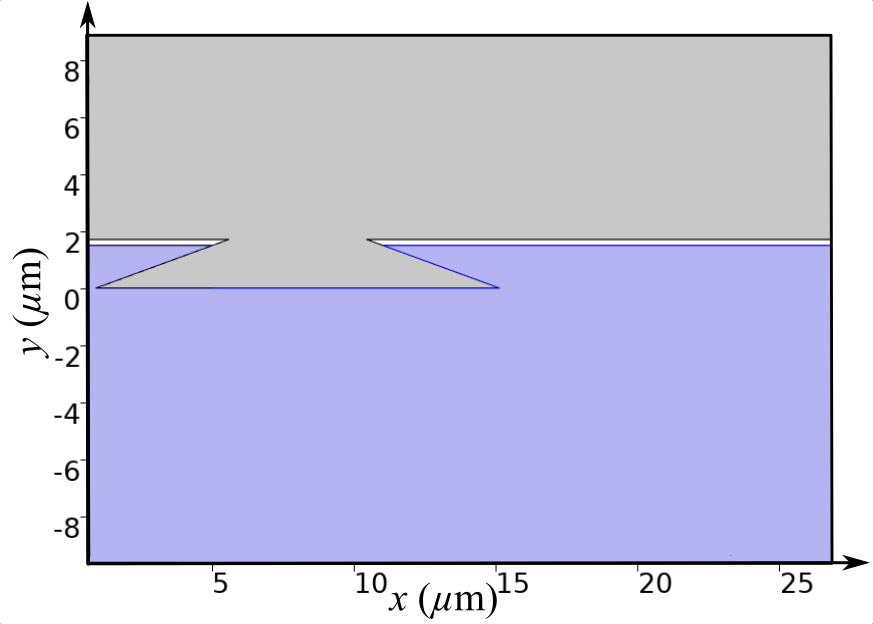}} \hspace{0.0003cm}
\subfloat[]{\includegraphics[scale=0.33]{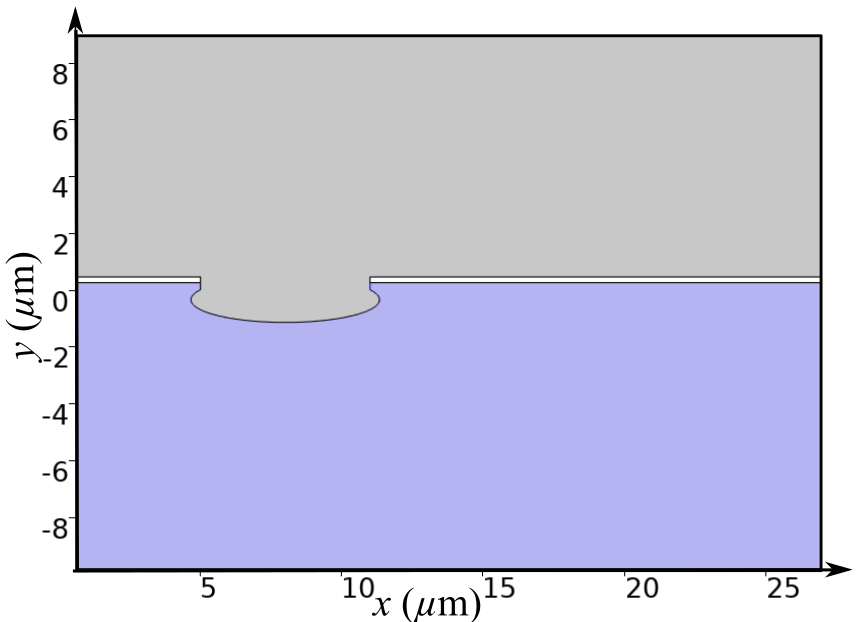}}

	\caption{
	Cross-sections of the modeled CPW resonators which are designed to exhibit increased participation at (a) in the silicon substrate (blue color), at (b) substrate-to-air, (c) metal-to-air, and (d) metal-to-substrate interfaces. Grey color denotes vacuum and white the superconductor.}\label{fig:iso-aniso-geoms}
\end{figure}

Along the lines of reference~\cite{Woods}, we model four CPW cross-sections that together produce condition numbers falling below $10^4$ for $\mathrm{P}$ and $\mathrm{F}$. These cross-sections, combining isotropic and anisotropic etching, are shown in \fref{fig:iso-aniso-geoms}. Each of these cross-sectional geometries is designed to increase the participation ratio of a single lossy region in comparison to the other regions and samples. The samples increasing the participation of the SA and MS interfaces and of silicon utilize isotropic etching into the substrate to produce the desired participation ratio characteristics. Mathematically, Bezier curves are utilized to obtain the smooth shapes in the models. The sample focusing the participation into the MA interface is chosen based on our initial geometric-dependence studies of section~3. Due to the way Bezier curves are used for creating the smooth shapes in the CPW gaps, the thickness of the SA interface layer varies slightly, but is approximately constant 1.5~nm when the thickness is mainly in the $y$-direction and 5~nm when the thickness is mainly in the $x$-direction, 
whereas the rest of the interface layers are exactly 5~nm thick. With these samples and dielectric constant values of 10 for the interface regions, one obtains a participation matrix
\begin{equation} 
\mathrm{P} = \frac{1}{100}
\begin{blockarray}{ccccc}
\mathrm{MS} & \mathrm{SA} & \mathrm{MA} & \mathrm{Si} \\
\begin{block}{(cccc)c}
0.030 & 0.090 & 0.0064 & 58.0 & \text{Sample SA} \\
0.080 & 0.058 & 0.0028 & 87.0 & \text{Sample MS} \\
0.046 & 0.099 & 0.049 & 65.0 & \text{Sample MA} \\ 
0.062 & 0.0096 & 0.0015 & 89.0 & \text{Sample Si} \\
\end{block},
\end{blockarray}
\end{equation}
which operates on the loss tangent vector to produce the reciprocals of quality factors as described by equation~\eqref{Q_TLS-matrix}. For this matrix $\kappa\left(\mathrm{P} \right) \approx 7800$. 

\subsection{Results}

The optimization problems are solved using the interior-point optimization (IPOPT) algorithm \cite{ipopt} implemented in the \texttt{fmincon} optimization tool of \texttt{MATLAB}~\cite{Matlab}.

Since these CPW resonators are not physically fabricated, we use simulations to obtain representative measurement data for the optimization. We computed the effective dielectric constants and quality factors from the field solutions using equations~\eqref{eps_eff-constraint-matrix} and~\eqref{Q_TLS-matrix}. Then, normally distributed noise is added to the quality factor and resonance frequency data using \texttt{MATLAB}'s \texttt{randn} function to mimic actual experiments. When creating the data, the values listed in the row ``Direct'' in table~\ref{table:computedValues} are used. For all the samples, the resonator length is set to $l = 30$~mm. However, since our modeling domains are two-dimensional, the length is merely a scaling factor for the resonance frequencies and may be chosen arbitrarily.

To monitor how well we can reproduce the dielectric constants of the lossy interfaces, we assumed  $\epsilon_\mathrm{Si}$ fixed at $11.6$, by bounding it from above and below in the optimization.\footnote{This is not to say we could not solve for $\epsilon_\mathrm{Si}$ simultaneously with the others. Note, however, that as the Si filling factors are also considerably larger than those of the interfaces, $\epsilon_\mathrm{eff}$ is more sensitive to changes in $\epsilon_\mathrm{Si}$.} This is justified in the sense that the dielectric constant of the substrate is usually the one known with the least uncertainty (except for that of vacuum). With sufficiently low noise levels, both approaches yield satisfactory results. Hence, in this case study, we keep the noise related to the measurements high enough to observe differences in these approaches, yet at a reasonably low level to study whether we can still reproduce the noiseless values: table~\ref{table:measurementData} describes the data we use as those for the measurements as opposed to the originally obtained simulation data. Identical data are used in the series and parallel approaches to ensure a fair comparison. Moreover, identical applicable solver settings, initial values, and tolerances are utilized in both cases.

\begin{table}[h!]
	\begin{center} 
		\caption{Resonance frequencies and quality factors used as measurement data for the inverse problems. Here, M refers to values used as measurement data, and D refers to those obtained directly from simulations, i.e., without added noise. The values are compared with the predictions obtained from the inverse problem solutions. S refers to series approach and P refers to parallel approach. The D columns show the absolute values, whereas the rest of the columns denote deviations from those. The samples are in the order of rising resonance frequency. Each ``Sample $i$'' refers to the sample designed to exhibit pronounced participation in the material region $\Omega_i$. In the parallel approach, the constraints were satisfied within tolerance $\tau = 7 \times 10^{-4}$.} 
		\label{table:measurementData} \small \footnotesize
		\begin{tabular}{l|c|c|c|c|c|c|c|c| }

			\multicolumn{1}{c}{} & \multicolumn{4}{c|}{ $f_0$ [GHz]} & \multicolumn{4}{c}{$Q_\mathrm{TLS}$ $[\times 10^6]$} \\
			\multicolumn{1}{c}{} & D & M & S & P & D & M & S & P \\
			\hline
			\text{1. Sample Si} & $2.1173$ & $+0.0004$ & $+0.0008$ & $+0.0000$ & $1.9136$ & $+0.0134$ & $+0.0130$  & $+0.0209$ \\ \hline
			\text{2. Sample MS} & $2.2723$ & $+0.0014$ & $+0.0004$ & $+0.0000$ & $1.1857$ & $+0.0459$ & $+0.0458$ & $+0.0275$  \\ \hline
			\text{3. Sample MA} & $3.1762$ & $-0.0018$ & $+0.0000$ & $+0.0004$ & $0.3379$ & $-0.0564$ & $-0.0304$ & $-0.0394$ \\ \hline
			\text{4. Sample SA} & $3.4104$ & $+0.0007$ & $-0.0001$ & $+0.0001$ & $1.1925$ & $+0.0215$ & $+0.0186$ & $+0.0278$ \\
		\end{tabular}
	
	\end{center}
\end{table}

\begin{table}[h!]
	\begin{center} 
		\caption{Utilized and computationally obtained material parameters. Direct approach refers to the parameters used in the simulations for computationally creating representative measurement data for the inverse problems. These values are similar to those used and obtained in reference~\cite{Calusine}. Series and parallel approaches refer to those obtained as solutions to the inverse problems. The direct approach row shows the absolute values, whereas the rest of the rows denote deviations from those. For Series$^*$ the noise amplitude for $f_0$ was reduced with a factor of 1/10 and for Series$^{**}$ with a factor of 1/100, otherwise utilizing the same data. Note that $\epsilon_\mathrm{Si}$ was bounded from above and below to yield exactly 11.6.}
		\label{table:computedValues} \footnotesize
\begin{tabular}{c|c|c|c|c|c|c|c|c|}

			Approach & 
			$\epsilon_\mathrm{MS}$ & $\epsilon_\mathrm{SA}$ & $\epsilon_\mathrm{MA}$ & $\epsilon_\mathrm{Si}$ & $\mathrm{tan}(\delta_\mathrm{MS})$ & $\mathrm{tan}(\delta_\mathrm{SA})$ & $\mathrm{tan}(\delta_\mathrm{MA})$ & $\mathrm{tan}(\delta_\mathrm{Si})$ \\ \hline 
			
			Direct & $ 11.6 $ & $5$ & $10$ & $11.6$ & $ 5.9 \times 10^{-4}$ & $ 7.1 \times 10^{-4}$ & $ 3.9 \times 10^{-3}$ & $1.2 \times 10^{-7}$ \\ \hline
			Parallel & $ +3.4 $ & $-0.4$ & $+5.0$ & $+0.0$ & $ +1.7 \times 10^{-4}$ & $ +0.3 \times 10^{-4}$ & $ +1.1 \times 10^{-3}$ & $+0.1 \times 10^{-7}$ \\ \hline
			Series & $-6.2$ & $+0.5$ & $-3.8$ & $+0.0$ & $-2.7 \times 10^{-4}$ & $-4.0 \times 10^{-4}$ & $+1.1 \times 10^{-3}$ & $-1.1 \times 10^{-7}$ \\ \hline
			Series$^*$ & $-1.4$ & $+3.9$ & $+4.5$ & $+0.0$ & $-1.1 \times 10^{-4}$ & $-3.6 \times 10^{-4}$ & $+2.5 \times 10^{-3}$ & $+0.2 \times 10^{-7}$ \\ \hline
			Series$^{**}$ & $+0.3$ & $+1.1$ & $+3.0$ & $+0.0$ & $-0.1 \times 10^{-4}$ & $-2.3 \times 10^{-4}$ & $+2.1 \times 10^{-3}$ & $+0.0 \times 10^{-7}$ \\

		\end{tabular}

	\end{center}
\end{table}

The results are shown in table~\ref{table:measurementData} and table~\ref{table:computedValues}. Both approaches reproduce the quality factor and resonance frequency data with small relative error. The most significant difference in the predictions is that the parallel approach reproduces the noiseless $f_0$ data almost perfectly, while the series approach shows larger yet still minor deviation. The resonance frequency prediction of the parallel approach deviates from the noiseless data at most 0.4~MHz, while the maximum deviation in the predictions of the series approach is 0.8~MHz.

Comparing the data in table~\ref{table:computedValues}, one notices that the parallel approach performs better in terms of predicting the dielectric constants and loss tangents. Whereas the parallel approach overestimates $\epsilon_\mathrm{MS}$ and $\epsilon_\mathrm{MA}$, the series approach underestimates them: especially $\epsilon_\mathrm{MS}$ is crudely underestimated by the series approach and $\epsilon_\mathrm{MA}$ overestimated by the parallel approach. Both approaches give a decent estimate for $\epsilon_\mathrm{SA}$. The maximum absolute deviation from the dielectric constant values utilized to obtain the noiseless data is 6.2 for the series approach and 5.0 for the parallel approach. Sum of the deviations in the dielectric constants for the series approach is 10.5 and 8.8 for the parallel approach. The fact that $\epsilon_\mathrm{MS}$ and $\epsilon_\mathrm{MA}$ are much larger than $\epsilon_\mathrm{SA}$ is only conserved in the parallel approach. In addition to the more accurate prediction of the dielectric constants by the parallel approach, it yields also a much closer match for the loss tangents: especially the substrate loss tangent is crudely underestimated by the series approach, whereas the parallel approach predicts it almost perfectly. In both approaches, the relatively low quality factor in the noisy case for the sample exhibiting increased participation in the MA interface results in overestimation of $\mathrm{tan}(\delta_\mathrm{MA})$. The predictions obtained using the series approach seem to be very sensitive to noise in the initial data: the effect of the variations in the interfacial dielectric constants on resonance frequencies is much smaller than their effect on losses. Hence, it may be essential to optimize the dielectric constants against both $Q_\mathrm{TLS}$ and $f_0$ measurement data.

There are at least two reasons for the reasonable reproduction of the quality factor data by the series approach despite the discrepancies in the loss tangents. Firstly, with the estimated values of $\mathrm{tan}(\delta_\mathrm{Si})$, the losses at the interfaces, the loss tangents of which are of the order of $10^{-4}$ to $10^{-3}$, still dominate the quality factors. Thus the prediction of the quality factors is not very sensitive to the observed variation in $\mathrm{tan}(\delta_\mathrm{Si})$. Secondly, the participation matrices differ significantly between these cases because of the different predictions for the dielectric constants. For example, the underestimation of $\epsilon_\mathrm{MS}$ by the series approach leads to overestimation of $p_\mathrm{MS}$, in line with equation~\eqref{SAScale}. Consequently to match the quality factor data, a lower $\mathrm{tan}(\delta_\mathrm{MS})$ is predicted.

These results highlight that the series approach is especially prone to errors arising from noisy resonance frequency data, and erroneous predictions of dielectric constants propagate to erroneous loss tangent predictions. Consequently, within these noise levels, only the parallel approach yields results accurate enough to serve as a starting point for further optimization of the cross-sectional geometries of this set of samples.

In contrast, utilizing the same parameters in this case study, but decreasing the amplitude of noise in the resonance frequencies with a factor of 1/100, the series approach yields very good results. Decreasing the amplitude with only a factor of 1/10, the results are not quite as accurate, but already comparable to the results of the parallel approach for noisier data. This supports using the simpler series approach instead of the parallel approach for such resonance frequency noise levels. 

\subsection{Discussion: predictions for quality factor}

The samples utilizing isotropic etching in this case study reach quality factors between $10^6$ and $2 \times 10^6$. These relatively wide or deeply trenched structures guarantee reduced participation at the interfaces, in line with the trends of the geometric dependence observed in section~3. Moreover, the utilization of isotropic etching gives the trenched gap a rounder shape, eliminating sharp angles, such as those in \fref{fig:fieldProfiles}. The fact that the sample~MA, on the other hand, exhibits a much lower quality factor is not very surprising since it is designed to yield a high MA participation and the MA was chosen to have the highest loss tangent. Of course, to optimize the cross-sectional geometries in terms of the quality factors resulting from a given fabrication process, systematic measurements with enough instances to obtain statistically meaningful results are required to find the loss tangents and dielectric constants of the different materials present in the devices.

\section{Summary and conclusions}

Dielectric loss due to TLS excitations is a typical hindering factor for the operation of cQED devices. In particular, such losses in the thin dielectric layers at material interfaces and in the silicon substrate may limit the quality factors of CPW resonators, essential in cQED.

We modeled the dielectric losses in CPW resonators using finite-element modeling by computing the electric-field participation ratios of the dielectric regions in such devices. We considered the effects of the geometric features of the CPW cross-section and of the dielectric constants of the different regions on the loss, and consequently, on the TLS-limited quality factors of the CPW resonators. The results support the view that increasing the physical footprint of the devices by trenching into the substrate and widening the structure mitigates dielectric losses. Furthermore, our systematic study indicates that designing optimally sloped sidewalls in the gap between the center conductor and the ground plane of the CPW can play a role in reaching high quality factors. Our results also show that variations in the dielectric constants of the lossy interface regions can significantly affect the predictions for the dielectric losses. The MA interface participation is particularly sensitive to uncertainties in the dielectric constants of the other interfaces. For example, our results indicate an 87\% difference in $p_\mathrm{MA}$ between $\epsilon_\mathrm{MS} = 5$ and  $\epsilon_\mathrm{MS} = 15$ for two otherwise identical samples.

To minimize the electric-field participation in the lossy regions of the resonators, it is essential to know the dielectric constants and loss tangents of the dielectrics with reasonable accuracy. We presented a nonlinear optimization method to find these quantities from quality factor and resonance frequency measurements. We considered two approaches, the series approach and the parallel approach to solve this inverse problem. In the series approach, the dielectric constants are first solved from resonance frequency data, and this solution is then used to solve the loss tangents using the measured quality factors. In the parallel approach, one finds the dielectric constants and the loss tangents simultaneously, minimizing the error with respect to quality factor data, whereas the resonance frequency data gives rise to a nonlinear constraint for the dielectric constants. Both approaches yielded reasonable estimates for the measured quality factors and resonance frequencies with reasonably noisy data, but the parallel approach was more accurate in terms of finding the right dielectric constants and loss tangents. In conclusion, for data with significant noise, e.g., with noise amplitude of 1~MHz for frequencies of the order of 1 to 5~GHz in the measured resonance frequencies, it is important to match the resonance frequency and quality factor data simultaneously to obtain reasonable estimates for dielectric constants and loss tangents. If the noise amplitude in the resonance frequency data is approximately 1/10th of this or less, the series approach is preferable due to its simplicity. 

Utilization of our inverse-problem approach requires a set of CPW resonators that result in well-conditioned participation and filling factor matrices. In general, the condition numbers of these matrices for a set of CPWs can be as high as of the order of $10^6$, if attention is not paid to this issue. In such cases, the uncertainties in the predictions for the loss tangent and for the dielectric constant are excessively high, and consequently the ability to predict dielectric losses and design low-loss cross-sectional CPW geometries is crippled. In line with reference~\cite{Woods}, we designed a set of CPW resonator cross-sections that yield condition numbers falling below $10^4$ for the matrices, allowing us to find the loss tangents and dielectric constants with a reasonable accuracy. Moreover, given realistic loss tangent values, quality factors exceeding $2 \times 10^6$ may be attained, as shown previously, e.g., in references~\cite{Calusine, Woods}. Note also that as different fabrication techniques can result in different interface qualities, samples fabricated in the same batch should be analyzed when utilizing the presented techniques.

Our nonlinear optimization approach complements and is in contrast to recent works, where, e.g., experimentally feasible values for loss tangents and dielectric constants have been assumed~\cite{Wenner}, or approximate scaling laws have been utilized to account for the variation in dielectric constants~\cite{Calusine, Woods}. Even though the measurement data we used to benchmark our approaches were fictional in the sense that they were created by adding noise to simulation data, this scheme represents a realistic model for possible experiments and importantly, it allowed us to conveniently investigate the robustness of our approaches against different noise levels. According to the results, our methodology is very promising in resolving the material parameters of the devices and provides new tools for driving improvements in superconducting CPW resonators. Thus it is a step toward more reliable modeling, and consequently more accurate design, of low-loss cQED devices.

Accurate measurement data are required to further validate and develop our approach. To minimize the resulting uncertainties, multiple measurements on nominally identical devices should be carried out. To extend the methodology to situations where coupling to the transmission line is significant, appropriate correction terms should be considered~\cite{Besedin}. Furthermore, our approach would benefit from highly accurate cross-sectional scans, since the thicknesses of the lossy interface layers play also a role in the prediction of the dielectric losses. Finally, we note that other loss mechanisms in cQED devices should be investigated as well: the total loss in a device with negligible dielectric loss may be dominated by, e.g., losses related to vortices \cite{Song, Bothner, Chiaro}, radiation \cite{Sage, Sandberg}, or quasi-particle excitations \cite{Sun, Wang14, de Visser, Nsanzineza}.

\section*{Acknowledgment}

The authors acknowledge support from the Academy of Finland through its Centre of Excellence in Quantum Technology (QTF) (Grant No. 312300), the QUESS project funded by the European Research Council (Grant No. 681311), and the Kvanttitietokone project funded by Jane and Aatos Erkko Foundation and the Technology Industries of Finland Centennial Foundation.

\end{document}